\documentclass{svjour3}                     
\smartqed  
\pdfoutput=1
\usepackage{graphicx}
\usepackage{natbib}

\usepackage{type1cm} 
\usepackage{url} 
\usepackage{subfigure}
\usepackage{verbatim}
\usepackage{amsmath,amssymb}
\usepackage{tikz}
\usetikzlibrary{arrows,decorations.pathmorphing,backgrounds,positioning,fit}
\usepackage{rotating}
\usepackage{cleveref}
\usepackage{xspace}
\usepackage{numprint}

\bibpunct{(}{)}{;}{a}{,}{,}

\newcommand{\alg}{MOSES\xspace}
\newcommand{\Alg}{\alg algorithm\xspace}

\newcommand{\todo}[1]{(\textcolor{blue}{\textit{TODO:}} #1)}
\newcommand{\sg}[0]{\ensuremath{s_Z}}

\newcommand{\boldrotate}[1]{\begin{sideways}\textbf{#1}\end{sideways}}

\newcommand{\CZ}{\mathcal{C}_{\mathrm{z}}}

\graphicspath{{pdffigs/}{epsfigs/}{mpsfigs/}}

\journalname{Social Network Analysis and Mining}
\begin{document}

\title{
Using Model-based Overlapping Seed Expansion to detect highly overlapping community structure.
\thanks{This research was supported by Science Foundation Ireland (SFI) Grant No. 08/SRC/I1407.}
}

\author{Aaron F. McDaid         \and    Neil J. Hurley 
}

\institute{Aaron McDaid \at
              Clique Research Cluster, UCD Dublin, Ireland \\
              \email{aaronmcdaid@gmail.com}           
           \and
           Neil Hurley \at
              Clique Research Cluster, UCD Dublin, Ireland \\
              \email{neil.hurley@ucd.ie}           
}

\date{Received: date / Accepted: date}

\maketitle

\begin{abstract}
As research into community finding in social networks progresses, there is a need for algorithms capable
of detecting overlapping community structure. Many algorithms have been proposed in recent years that are
capable of assigning each node to more than a single community.
The performance of these algorithms tends to degrade when the ground-truth contains a more highly
overlapping community structure, with nodes assigned to more than two communities.
Such highly overlapping structure is likely to exist in many social networks,
such as Facebook friendship networks.
In this paper we present a scalable algorithm, \alg, based on a statistical model of community structure,
which is capable of detecting highly overlapping community structure, especially when there is variance
in the number of communities each node is in.
In evaluation on synthetic data  \alg\ is found to be superior to existing algorithms, especially
at high levels of overlap. We demonstrate \alg\ on real social network data by analyzing the networks of friendship links between
students of five US universities.
\end{abstract}



\section{Introduction}
\label{sec:intro}

In this paper we introduce \alg{}, a Model-based Overlapping Seed ExpanSion\footnote{Our C++ implementation of \alg{} is available at \url{http://sites.google.com/site/aaronmcdaid/moses}.}
algorithm, for finding overlapping communities in a graph.  The algorithm is designed to work well in applications, such as social network analysis, in which the graph is expected to have a complex, highly-overlapping community structure.

\begin{figure}[t]
\centering
\includegraphics[scale=0.25]{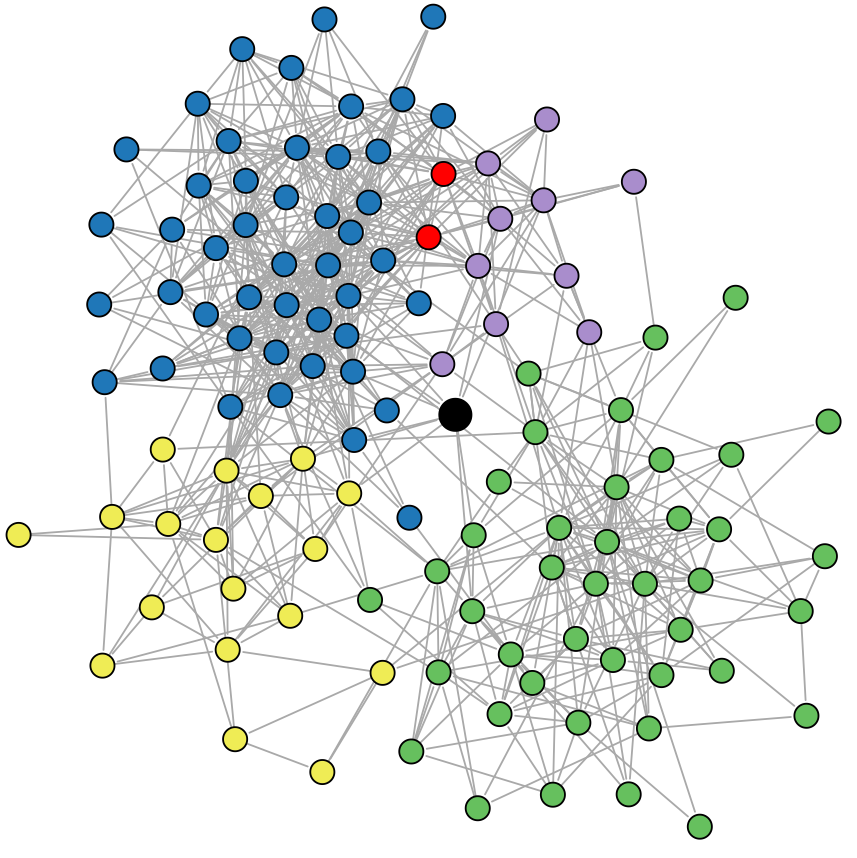}
\caption{\label{Georgetown4240}Four communities of a single user (the node in black) of Facebook, as determined by \alg{}. Two other users (red) have been assigned to both the blue and purple communities.
A typical user, like many of those in this diagram, will be a member of several communities, which we have not attempted to visualize here. See \cref{sec:fb} }

\end{figure}

Many of the algorithms for finding communities in graphs have been limited to \emph{partitionings}, where each node
is assigned to exactly one community. While there are still very many open questions about the basic structure of
empirical graphs, it is difficult to accept that a partition is an appropriate description of the complete community
structure in a graph. \cite{PartitioningHarmful} show that partitions will break many large cliques in
empirical networks, and hence we cannot assume that partitioning will preserve much community structure.

In recent years, many algorithms have been proposed to detect overlapping communities. We repeat experiments similar to
those carried out in \cite{GCE}, which show that many such algorithms are only capable of detecting weakly overlapping community structure,
where a typical node is in just two communities. If we are to be able to make reasonable inferences about the
community structure in empirical graphs, we need algorithms capable of detecting highly overlapping communities, if only
so that we can credibly rule out highly overlapping community structure for a given graph.

\cite{LeskovecNCP} claim that large scale community structure may not exist in typical empirical graphs, by showing
that it is difficult or impossible to find subgraphs with good \emph{conductance}, a measure comparing the number
of edges inside a cluster to the number of edges which travel from inside to the outside of a cluster. However,
in this paper we will show that such structure may indeed exist, and be detectable, even when the \emph{conductance}
suggests otherwise.

The method presented here is similar in spirit to many existing algorithms, in that a global objective function is defined
to assign a score to each proposed community assignment. The algorithm proceeds by using simple heuristics to search for
communities in the graph, greedily finding a (local) maximum of our objective function. This allows for scalability as the \alg{} objective function
can be efficiently updated throughout.

\subsection{Structure of this paper}
We first briefly consider related work in the field on overlapping community finding. Then, in \cref{TheMOSESModel,sec:methods} we
introduce our objective function and describe the algorithm.

Next, we consider the \emph{network community profile} of \cite{LeskovecNCP} and show, perhaps surprisingly, that
large values of conductance are not incompatible with the existence of strong, easily detectable, community structure.
We conclude with an analysis of a Facebook friendship network from five US universities.

Then, there is an evaluation of the algorithm with two types of synthetic benchmark data, the LFR benchmarks proposed in \cite{fortunato-2008}, and a second model that allows for greater variance in the community overlap structure.

\paragraph{Notation}  In this paper we consider the community assignment problem on an unweighted, undirected graph $G$, with vertices $V$ and edges $E$ and no self-loops.
Boldcase letters, such as $\mathbf{Z}$, $\mathbf{z}$ denote column vectors with the uppercase $\mathbf{Z}$ referring to a random vector
variable and the lowercase $\mathbf{z}$ referring to a particular realization of $\mathbf{Z}$.
We use capital Roman letters, such as $\mathrm{Z}$ to denote random matrices and their realizations.
The components of random matrices are denoted by the corresponding uppercase letter e.g. $Z_{ij}$, while the components of matrix realizations are denoted by the corresponding lowercase letter e.g. $z_{ij}$.
The notation used in the description of  the  \alg{} model is summarized in \cref{table:defs}.

\begin{table}[hbtp]

\centering
\caption{\label{table:defs}Basic definitions. In this table, $(0,1)$ means a real number between zero and one. $\{0,1\}$ means the set with just two elements, zero and one.}
\begin{tabular}{|l|l|p{7.5cm}|}

	\hline
	 & \textbf{Range} & \textbf{Description} \\
	 \hline \hline

	$N$                     &  $\mathbb{N} $                      &  Number of vertices in $G$.           \\
	$X_{ij}, x_{ij}$        &  $\{0,1\}^{N \times N} $              &  Adjacency matrix of a simple, undirected, unweighted graph, $G$. $X_{i,i}=0$, $X_{ij} = X_{ji}$          \\

	$Q$                     &  $\mathbb{N} $                      &  Number of communities. Sometimes called $K$ in related work.           \\
	$\alpha     $           & $ (0,1)^Q $                           &  Vector of length $Q$ giving the memberships proportions. For partitioning, $\sum_{1 \leq i \leq Q}\alpha_i = 1 $ \\
	$Z_{iq}, z_{iq}$        &  $\{0,1\}^{N \times N} $              & Community assignment matrix, one  if node $i$ in community $q$, zero otherwise.           \\

	$\pi     $              &  $(0,1)^{Q \times Q}$                &  Connection probabilities between pairs of clusters. Most models use a different or simplified form. In \alg{}, $p_{in}$ and $p_o$ take the place of $\pi$.  \\
	\hline
	\multicolumn{3}{|l|}{ \alg{}-specific:} \\
	\hline

	$p_o$                   &  $(0,1)$                 &          Probability that two nodes  connect, independent of community structure.              \\
	$p_{in}$                &  $(0,1)$                 &  Probability that two nodes connect, due to their assignment to a common community.               \\
	$Q_z$                   &  $\mathbb{N}$        &  Number of non-empty communities observed in a community assignment $\CZ$.         \\
	$\CZ$                   &          &  A community assignment corresponding to an assignment matrix $\mathrm{z}$.        \\
	$m$                     & $\mathbb{N}$ & Mean number of communities in $G$.\\
	$n_q$                   & $\mathbb{N}$ & The number of nodes in community $q$. It is a function of $\mathrm{z}$.\\
	$s_z(i,j)$              & $\mathbb{N}$ &  Number of communities in $\mathrm{Z}$ shared between node $i$ and node $j$.         \\
	\hline
\end{tabular}
\end{table}

\section{Related Work}
\label{sec:related}

While there is no single generally accepted definition of a community within a social network, most definitions try to encapsulate the concept as a sub-graph that has few external connections to nodes outside the sub-graph, relative to its number of internal connections. We find the following distinctions useful in characterizing commonly-used community definitions:
\begin{enumerate}
\item \emph{Structural communities}: A deterministic set of properties or constraints that a sub-graph must satisfy in order to be considered a community is given and thus a decision can be made on whether any particular sub-graph is, or is not, a community, e.g. we may consider all maximal cliques to be communities.  Thus finding such communities is a process of searching the graph for all sub-graph instances that satisfy the defining properties.
\item \emph{Evaluated communities}: Every sub-graph is considered to be a community to a certain extent, given by the value of a \emph{community fitness function}.  The fitness function may be local or global in nature and sometimes is associated with the entire community decomposition rather than with each single community. 
\item \emph{Algorithmic communities}:  As pointed out in \cite{fortunato-2010}, often there is no explicit definition of a community, other than as the sub-graphs that result from some community extraction algorithm. A good example of this is the edge-betweenness algorithm of \cite{newman-2004-2}.
\end{enumerate}
The last decade has seen a lot of publications on the topic of community detection in networks.  For a good review, see \cite{fortunato-2010}.  Much work has concentrated on \emph{modularity} maximization algorithms, that produce partitions of graphs in which each node is assigned to a single community. Modularity defines evaluated communities, where the community fitness is related to its number of internal edges relative to its expected number in a particular `null model'.   While modularity maximization results in a decomposition of the entire network into a partition of communities, in fact, a more general view of community-finding is from the node perspective as \emph{community assignment}, i.e. the task is to assign each node in the graph to the communities (if any) it belongs to and we may describe algorithms for community-finding as \emph{community assignment algorithms} (CAAs).

A number of CAAs that allow overlapping communities have emerged since 2005 \cite{GCE, palla-2005, clauset-2005,
gregory-2007,gregory-2009:1,gregory-copra, mishral-2007,lancichinetti-2009,
baumes-2005,shen-2009,AhnEdgePartition}.  For example GCE \cite{GCE}, LFM \cite{lancichinetti-2009} and Iterative Scan (IS) \cite{baumes-2005} find evaluated communities.  Each uses various local iterative methods to expand (or shrink) proposed communities such that some function of the density of the communities is maximized, but the decision on whether a proposed community is accepted or not depends on somewhat arbitrary criteria.  At the other end of the spectrum, the Clique Percolation Method (CPM) of \cite{palla-2005} has proved very influential and is essentially a structural community-finding algorithm, where communities are defined as sub-graphs consisting of a set of connected $k$-cliques. 

With the recent release of the LFR synthetic benchmark graphs \cite{lancichinetti-2009:2}, it has become possible to more thoroughly explore the performance of these different approaches.
Studies on this benchmark data have illustrated that performance of the algorithms generally degrades
as nodes are shared between larger numbers of communities \cite{GCE}.
It is our contention that real-world social communities can in fact contain rich overlapping structures like those of the overlapping LFR benchmarks and that it is necessary to develop CAAs that perform well when on average each node is assigned to multiple communities.
There is need for further extensions of these synthetic benchmarks as, for example, the current LFR model places each overlapping node into exactly the same number of communities.

Model-based CAAs have the advantage of being based on a model which can explain the rationale of the communities found,  thus avoiding the often arbitrary criteria which are used in many overlapping CAAs.
We develop a scalable, model-based CAA that performs well on highly overlapping community structures.  In the next section, we review the model-based network algorithms that are most relevant to our approach.

\subsection{Model-Based Community-Finding}
In model-based community-finding, the graph $G$ is considered to be a realization of a statistical model.  Assuming unweighted, undirected graphs, with no self-loops, the graph edges are represented by a random symmetric adjacency matrix $\mathrm{X}$ such that $x_{ij} = x_{ji} = 1$ if an edge connecting nodes $i$ and $j$ exists and zero otherwise.  Statistical network models are reviewed in \cite{Goldenberg-09}.   Of particular interest in the context of the work presented here is the \emph{stochastic blockmodel} introduced in \cite{Nowicki-01} which is also referred to as the \emph{Erdos-Renyi Mixture Model for Graphs} (ERMG) in \cite{Daudin-08}. We will use our our notation, as defined in \cref{table:defs}, when describing the related work.

The ERMG assumes a partitioning of the graph into communities, so that community assignments can be described by the vector $\mathbf{z}=(z_1,\dots, z_N)^T$, where $z_i=q$ if node  $i$ is assigned to community $q$.  
The graph edges are assumed independent given the node assignments $\mathbf{z}$, and drawn from a Bernoulli distribution with connection probability dependent on the community assignments of the end-points:
\[
	P(X_{ij} = 1 | \mathbf{Z} = \mathbf{z}) = \pi(z_i, z_j) \equiv \pi_{z_i z_j}\,.
\]
Assuming that $\pi_{qr} = \pi_{rq}$, this leads to the conditional probability for $\mathrm{X}$ given $\mathbf{Z}$, 
\begin{eqnarray}
\label{eqn:ermg}
   P(\mathrm{X} | \mathbf{Z}, \mathrm{\Pi}) =\prod_{i=1}^N\prod_{j=i+1}^N \pi_{z_iz_j}^{x_{ij}} ( 1- \pi_{z_iz_j} ) ^{(1 - x_{ij})}\,,
\end{eqnarray}
where $\mathrm{\Pi}$ is the $Q \times Q$ matrix of inter-community connection probabilities $\{\pi_{qr}\}$. 
Each component of $Z$ is modelled as being a single draw from a multinomial $(1;\alpha)$;
where $\alpha$ is a vector of length $Q$ describing the memberships probabilities for each cluster.

Ultimately, the goal is to predict the unobserved community assignments $\mathbf{z}$.
In this section we will use \emph{parameter} to refer to quantities such as $\mathrm{\Pi}$ and $\alpha$
which describe connection probabilities and cluster membership proportions, and we will not refer to $\mathbf{z}$ as a parameter.
As discussed in  \cite{Nowicki-01} parameter estimation is difficult as the observed likelihood:
\[
	P(\mathrm{X} | \boldsymbol{\alpha}, \mathrm{\Pi}) = \sum_{\mathbf{z} \in \{1,\dots,Q\}^N} P(\mathrm{X},\mathbf{Z}=\mathbf{z} |  \mathrm{\Pi}, \alpha)
\]
cannot be simplified
and the Expectation Maximization (EM) algorithm requires, among other things, the conditional $\mathrm{P}(\mathbf{Z} | \mathrm{X} , \mathrm{\alpha}^{(t)} , \mathrm{\Pi}^{(t)} )$
when calculating the next estimates $\mathrm{\alpha}^{(t+1)}, \mathrm{\Pi}^{(t+1)}$, which is also intractable within these types of models.

In \cite{Daudin-08} a variational approach is taken to parameter estimation. In \cite{Zanghi-07} a heuristic algorithm is used to quickly attempt an approximate maximization of the complete-data log-likelihood, $\mathrm{P}(x, z | \alpha , \pi)$,
searching over $(\mathbf{z}, \mathbf{\pi}, \mathbf{\alpha})$ with $x$ fixed equal to the graph which has been observed.
An online estimation approach is used where the parameters, and cluster assignments, are incrementally updated using the current value of the parameters and new observations.  The algorithm is essentially a greedy maximization strategy.
The ERMG assumes a fixed number of communities. To decide between different values of $Q$, both \cite{Daudin-08} and \cite{Zanghi-07} use an Integrated Classification Likelihood (ICL) criterion to decide between competing models. 

\cite{NobileAllocationSampler} integrate out $Q$, creating a posterior density mass function defined over all clusterings, regardless of the number of clusters.
This means that model selection, such as the BIC and ICL, are unnecessary. This \emph{allocation sampler} is presented in terms of 
gaussian mixture models, but this technique is suitable in variety of contexts, including network modelling and for overlapping clusters.

The \alg{} model is similar to \cite{Nowicki-01}, in that the parameters such as $\alpha$ and $\pi$ are treated as nuisance parameters to be integrated out.
They do not integrate out $Q$. They propose a Gibbs sampler to sample from $(z, \alpha, \pi)$, effectively allowing them to numerically integrate out $\alpha$ and $\pi$.
\cite{NobileAllocationSampler} point out that this can often be analytically integrated out, allowing the algorithm to focus on estimating the quantities of interest,
which are typically the $z$ and $Q$.

\subsubsection{Overlapping Stochastic Block Modeling}
In \cite{Latouche:OSBM}, the standard ERMG is expanded to allow for overlapping communities and  the new model is named the Overlapping Stochastic Blockmodel (OSBM). 
Now the community assignments of a node $i$ may be described by a vector $\mathbf{Z}_i = ( Z_{i1},\dots, Z_{iQ})^T$, such that 
\[
	Z_{iq} = \left\{\begin{array}{cc}
	                                 1 & \mbox{node}\,\,i\,\,\mbox{in community}\,\, q\\
	                                 0 & \mbox{otherwise.}
	                                 \end{array} \right.
\]
The full latent structure may be described by the $N\times Q$ matrix $\mathrm{Z}$, with $i^{\rm th}$ column $\mathbf{Z}_i$.  As with the ERMG, it is assumed that  all the edges are independent, given $\mathrm{Z}$ and drawn from a Bernoulli distribution, with the probability $\pi(\mathbf{z}_i, \mathbf{z}_j)$ that an edge exists dependent on the (vector) community assignments $\mathbf{z}_i$ and $\mathbf{z}_j$ of its end-points, leading to a joint distribution of the same form (\ref{eqn:ermg}), with $\pi_{z_iz_j}$ replaced by $\pi(\mathbf{z}_i, \mathbf{z}_j)$.

The authors assume that the connection probabilities, $ \pi(\mathbf{z}_i, \mathbf{z}_j)$ can be written as sigmoid functions of a quadratic form $\mathbf{z}_i \mathrm{A} \mathbf{z}_j$ for a parameter matrix $\mathrm{A}$.
In a natural extension of the relationship between $z$ and $\alpha$ used in the (non-overlapping) block models, they choose a prior distribution on $\mathrm{Z}$ of the form:
\begin{equation}
\label{eqn:osbmprior}
	P(\mathrm{Z}|\boldsymbol{\alpha}) = \prod_{i=1}^N\prod_{q=1}^Q\alpha_q^{z_{iq}}(1-\alpha_q)^{1 - z_{iq}}\,,
\end{equation}
for parameters $\alpha_q \in [0,1]$.
The parameters of the model  are estimated using a variational strategy similar to that used in \cite{Daudin-08}.

While the models  of \cite{Nowicki-01} and \cite{Latouche:OSBM} allow for a large number of parameters, in practice, when evaluating on real datasets,  the parameter space is usually restricted to a much smaller number.  In \cite{Latouche:OSBM} for instance, this is done by considering restricted forms of the matrix $\mathrm{A}$, with only two free parameters.  The community-finding algorithm of \cite{Latouche:OSBM} is shown to out-perform the Clique Percolation Method of \cite{palla-2005} on synthetic data.


While our model is another form of overlapping SBM, our general approach
shares much in common with \cite{NobileAllocationSampler} as we have integrated out $Q$, the number of clusters, and $\alpha$ allowing our
algorithm to search over the space of all clusterings, regardless of the number of communities.
And our estimation method could be compared with \cite{Zanghi-07}, in that it is another method using a fast heuristic algorithm to greedily search over $\mathbf{z}$.


\section{The \alg{} Model}
\label{TheMOSESModel}
The model that drives \alg{} is essentially an OSBM but with some important differences to that presented in \cite{Latouche:OSBM}. In particular:

\begin{enumerate}
\item The connection probabilities $\pi(\mathbf{z}_i, \mathbf{z}_j)$ take a different form to those used in \cite{Latouche:OSBM};
\item The prior takes into account that community assignments that differ only by a relabeling of the communities are equivalent;
\item A distribution is placed on the number of communities $Q$, allowing $Q$  to be integrated from the prior, in the manner or \cite{NobileAllocationSampler}.
\end{enumerate}
We elaborate on these differences in the following:
\subsection{Connection Probabilities}
Let $\pi_{qr}\in[0,1]$ represent the probability that a node in community $q$ connects to a node in community $r$ and let $p_o$ denote a general underlying probability that nodes connect, independent of community structure.   Assume that these probabilities are all mutually independent. Hence, the probability that an edge does \emph{not} exist is given by:
\begin{eqnarray}
\label{eqn:mosesmodel}
	P(X_{ij} = 0|\mathrm{Z}, \mathrm{\Pi}) &=& 1- \pi(\mathbf{z}_i, \mathbf{z}_j) \\\nonumber
	                                       &= & (1-p_o)\prod_{q=1}^Q \prod_{r=q}^Q (1 - \pi_{qr})^{z_{iq} z_{jr}}\,.
\end{eqnarray}

In practice, we use $\mathrm{\Pi} = {\rm diag}(p_{in})$.  Thus, there is a single connection probability $p_{in}$ of within-community connections and there is no tendency for inter-community connections, other than the general tendency of nodes to connect represented by $p_o$. With this simplification, (\ref{eqn:mosesmodel}) becomes,
\[
	P(X_{ij} = 0|\mathrm{Z}, p_{in},p_o) = (1-p_o) (1 - p_{in})^{s_Z(i,j)}\,.
\]
where $s_Z(i,j)$ is a count of the number of communities assigned to both node $i$ and node $j$ in $\mathrm{Z}$. 

It is also possible to imagine a large community containing every node, which allows one to treat $p_o$ as being the internal connection probability of that community. This can then be used in the appropriate cell of an augmented $\mathrm{\Pi}$ matrix.

\subsection{Prior on $\mathrm{Z}$}
Assuming a uniform distribution on the parameters $\{\alpha_1,\dots\alpha_Q\}$ in (\ref{eqn:osbmprior}) and integrating over them, we obtain a prior of the form
\[
	P(\mathrm{Z} | Q) = \frac {1} {(N+1)^Q} \left(\prod_{q=1}^Q\frac {1} {{N \choose n_q}}\right) \,,
\]
where $n_q$ is the number of nodes assigned to community $q$. Furthermore, while there are $2^{N Q}$ possible values for $\mathrm{Z}$,  any permutation of the columns of $\mathrm{Z}$ results in the same community assignment, with just a different labeling on the communities. The $2^{NQ}$ possible matrices  can be partitioned into equivalence classes of matrices that differ only in a permutation of their columns. Let $c_{z}(Q)$ be the size of the equivalence class that $\mathrm{Z}$ belongs to.  Using $\CZ$ to denote the community assignment corresponding to the $c_{z}(Q)$ matrices in this equivalence class, we note that $P(\CZ | Q) = c_z(Q)P(\mathrm{Z}| Q)$.   Let $Q_z$ be the number of non-empty communities observed in $\mathrm{Z}$. If the actual number of communities is $Q_z+k$, then $\mathrm{Z}$ should contain $k$ columns of all zeros.  It follows that
\begin{equation}
\label{eqn:cz}
	c_{\mathrm{Z}}(Q_z+k) = {{Q_z+k} \choose k} c_{\mathrm{Z}}(Q_z)\,,
\end{equation}
since the $k$ communities with no nodes assigned to them must be allocated $k$ labels out of the $Q_z+k$ possible community labels.  Furthermore, 
\begin{equation}
\label{eqn:pz}
	P(\mathrm{Z} | Q_z+k) = \frac 1 {(N+1)^k} P(\mathrm{Z} | Q_z)\,.
\end{equation}
Now, choosing a \emph{Poisson} distribution for $Q$ with mean value $m$, using (\ref{eqn:cz})and (\ref{eqn:pz}),  and summing over $Q$ to obtain a prior on $\CZ$ that is independent of $Q$, we get 
 \begin{eqnarray}
 \label{eqn:pcz}
	P(\CZ ) &=& \sum_{k=0}^{\infty}P(\CZ | Q=Q_z+k)\frac {e^{-m}m^{Q_z+k}}{(Q_z+k)!}\\ \nonumber
	                              &=&  \frac {c_z(Q_z)} {(N+1)^{Q_z}}  \left(\prod_{q=1}^{Q_z}\frac 1{ { {N} \choose {n_q}}}\right)\frac{e^{-(\frac N {N+1})m} m^{Q_z}}{Q_z!} 
\end{eqnarray}
Finally,  if there are $p$ unique non-zero columns in $\mathrm{Z}$, which occur with multiplicity $o_1, \dots, o_p$, such that $Q_z = \sum_{k=1}^p o_k$, we note that $c_z(Q_z)$ is the multinomial coefficient: 
\[
       c_z(Q_z)  = \frac {Q_z!} {o_1! \dots o_p!}
 \]
 
 With (\ref{eqn:pcz}) and (\ref{eqn:ermg}), letting $\mathcal{L}(.) = \log P(.)$, it is now possible to write down the complete data log likelihood as
 \begin{equation}
 \label{eqn:obfunc}
 	F(\CZ,p_{in},p_{o}) = \mathcal{L}(\mathrm{X}| \mathrm{Z}, p_o, p_{in}) + \mathcal{L}(\CZ)\,.
\end{equation}
Strictly speaking, this might not be considered the \emph{complete} data, as we have integrated out $\alpha$ in a Bayesian manner.
For our purposes however, $\mathrm{z}, p_o, p_{in}$ will be referred to as the complete data.

As methods such as \cite{Daudin-08} that attempt to find the maximum likelihood estimators from the observed likelihood $\mathcal{L}(\mathrm{X})$ are too computationally expensive for large-scale networks,
and because we are more interested in estimating the clustering than in estimating the paremeters,
we follow an approach similar to \cite{Zanghi-07} and seek the $(\mathrm{\CZ}, p_{in}, p_{o})$ that maximizes (\ref{eqn:obfunc}).
Maximization of the complete data likelihood has been shown to result in good clusterings  in practice in the context of Gaussian mixture models. In the remainder of the paper, we will simply write $F(\CZ)$,
rather than $F(\CZ,p_{in},p_o)$, to emphasize our primary objective of finding an optimal $\CZ$.
 
We have integrated out $\alpha$, the cluster membership proportions.
If it was easy to analytically integrate out $p_{in}$ and $p_o$, giving us
$$
F^*(\CZ) = \int_0^1 \int_0^1 F(\CZ,p_{in},p_o) \mathrm{P}(p_{in}) \mathrm{P}(p_o) \mathrm{d}p_{in} \mathrm{d}p_o ,
$$
then this would allow us to consider $p_{in}$ and $p_o$ as nuisance parameters and to totally disregard them in our algorithm,
in the manner of \cite{NobileAllocationSampler}.
However,
it does not yet appear possible to do so. For convenience we chose to search for the triple $\widehat{(\CZ,p_{in},p_o)}$
that maximizes $F(\CZ)$.
Another alternative would be to sample these parameters in the manner of \cite{Nowicki-01}.

\section{The \alg{} Maximization Algorithm}
\label{sec:methods}

\tikzset{vertex/.style={circle,draw=gray!70,fill=gray!30} }
\tikzset{invertex/.style={vertex,black} }
\tikzset{frontiervertex/.style={invertex,blue!50} }
\tikzset{outvertex/.style={invertex,draw=gray,fill=white!0       }}
\tikzset{edge/.style={shorten <= 1pt, shorten >= 1pt,gray,thick} }
\tikzset{frontier edge/.style={edge, blue!50, dashed} }
\tikzset{out edge/.style={edge, dotted} }

\alg{}, similarly to algorithms based on modularity, is driven by a global objective function, $F(\CZ)$.  Except in the smallest of networks, it is not feasible to exhaustively search every possible community assignment,
calculating $F(\CZ)$ for each, and then remembering which got the best score.  In order to handle graphs with millions of edges,
we use a greedy maximization strategy in which communities are created and deleted, and nodes are added or removed from  communities, in a manner that leads to an increase in the objective function.

The change in the objective when an entire community is added or removed can be decomposed into a set of single node updates.
A single node update, adding it to, or removing it from, a community, changes $z_{iq}$ to $z'_{iq} = 1-z_{iq}$.
In order to avoid considering a node being connected to itself in the following expression,
which is not allowed in this model, we focus on the addition of a community in this discussion.
For convenience we define $\psi_{in} = 1-p_{in}$ and $\psi_{o} = 1-p_{o}$.

The objective, $F(\CZ)$, changes where node $i$ is being added to community $q$,
where $j$ iterates over the set of nodes already within $q$,
\begin{align*}
    \Delta F(\CZ)
				& = n_q \log \psi_{in}  - \sum_{ z_{jq}=1 } x_{ij} \log \psi_{in}              \\
    		& + \sum_{ z_{jq}=1 } x_{ij} \log\left( \frac {1 - \psi_o \psi_{in} {}^{s'_Z(i,j)}}{1 - \psi_o \psi_{in} {}^{s_Z(i,j)}} \right)     \\
		& + \log \frac1{\binom{N}{n_q'}} - \log \frac1{\binom{N}{n_q}}
\,,  
\end{align*}
where $s'_Z(i,j) = (-1)^{z_{iq}} + s_Z(i,j)$ is the number of common communities between $i$ and $j$ after the node update has taken place.
We note that we need the values of $s_Z$ only for those pairs of nodes that are connected, the edges.

The change in \emph{a priori} probability of $\CZ$, $\Delta\mathrm{P}(\CZ)$, is more complicated as it depends on whether the node update results in a change to $Q_z$, or not.
We estimate $m$, the mean value of $Q$ to be $\hat m = Q_z$, which allows us to simplify and approximate (\ref{eqn:pcz}) when considering small changes to $Q_z$. $m$ is fixed but unknown,
and hence $e^{-(\frac N {N+1})m}$  is a constant we can ignore for proportionality.
A small change in $Q_z$, increasing or decreasing it by 1, will make little change in the ratio $ \frac{ m^{Q_z}}{Q_z!} $, as $m$ has been estimated to approximate $Q_z$ .
\[
	P(\CZ ) \propto  \frac {c_z(Q_z)} {(N+1)^{Q_z}}  \left(\prod_{q=1}^{Q_z}\frac 1{ { {N} \choose {n_q}}}\right)
\]
Moreover, changes to $c_z(Q_z)$  depend on whether the node update results in a change to the number or multiplicity of unique columns in $\mathrm{Z}$.
In \alg{}, we assume that all the communities we have found are unique, estimating $c_z = Q_z!$.
This introduces an overestimate of the multinomial $c_z$,
and we would expect that this would introduce a bias towards finding duplicate communities. However, we have not yet observed a duplicate community in the
output of the algorithm. 

We use a combination of heuristics in an attempt to find good communities. These are edge-expansion, community-deletion and single-node fine-tuning. In the following, it is more useful to think of a community assignment $\CZ$ as a set of communities, with each community consisting of a set of nodes. We will use $\CZ \cup C$ for $C \subseteq V$ to denote the addition of a new community to $\CZ$.

%


\paragraph{Edge expansion}
In the initial phase of the algorithm, edges are selected at random from the graph and  a community is expanded around each selected edge in turn.
Initially the community consists of two nodes $C=\{v,w\}$. New nodes are added to $C$ from its \emph{frontier} i.e. the set of nodes not in $C$ but directly connected to nodes in $C$.
Nodes are added in a greedy manner, selecting the node $v^*$ in the frontier that maximizes $F(\CZ \cup \{ C\cup v\})$.
Expansion continues while the objective is the highest found so far.

When a proposed community is very small, its contribution to the objective may be negative even if it is a clique.
This is because, for a small community, $\mathrm{P}(\CZ)$ dominates $\mathrm{P}(X | \CZ, p_{in}, p_o)$ in $F(\CZ)$.
Hence, we use a small lookahead, whereby expansion of a community will continue, even if it would decrease the objective, unless $l$ consecutive expansions fail to raise the
objective.
In practice, we use $l=2$ and have found that large values of $l$ slow down the algorithm, without any significant improvement to the quality of the results.

Edges are chosen randomly with replacement to be subject to expansion.
Note that each subsequent time an edge is selected, it may expand into a different community, as, with each addition of a new community, the overlap counts $s_Z(i,j)$  change.
For the first community expansion $v^*$ is simply the node with most connections to $C$. Then,
as more expansions are performed, and more and more edges
are `claimed' by found communities, and $s_Z(i,j)$ increases, the expansion will favour edges with lower $s_Z(i,j)$.
Informally, we can say that $F(\CZ)$ favours finding communities of nodes which are densely connected by edges,
and that it has a preference for edges not already contained within other communities.

\paragraph{Community Deletion}
Periodically all the communities are scanned to see if the removal of an \emph{entire} community will result in a positive change in the objective.
This check occurs after each 10\% of the edges have been expanded and after the single-node fine tuning phase, so will happen 11 times.
The output of the algorithm will be the assignments after the last community deletion phase.
\[
	F(\CZ \setminus \{C\} ) > F(\CZ)
\]
\paragraph{Single-Node Fine Tuning}
The fine tuning phase takes place at the end of the edge expansion phase. It is inspired by the method of \citet{blondel-2008}.   In this phase, each node is examined in turn by removing it from
all the communities it is assigned to and then considering adding it to the communities to which it is connected by an edge. 
As always, the decision to insert a node into a neighbouring community depends on whether it results in a positive change to $F(\CZ)$. 

\paragraph{Estimating $p_{in}$ and $p_o$}
The \alg{} algorithm does not require the user to specify the two connection probability parameters. The algorithm estimates these itself as it proceeds.
Only one input, the graph, is supplied to the MOSES software.
It can be shown that, for a given $z$ and $x$, and as a function of $p_{in}$ and $p_o$, the value of $F(\CZ, p_{in}, p_o)$ depends
on simple summary quantities such as the frequency of various values of $s_Z(i,j)$ across the edges. This allows us to efficiently select the values of $p_{in}$ and $p_o$,
given the current estimate of the communities, which maximize $F(\CZ)$.

\section{Evaluation}
\label{sec:eval}

\subsection{Do empirical networks have highly overlapping community structure?}

\begin{figure}[htbp]
\centering
\includegraphics[scale=0.95,trim = 0mm 5mm 0mm 5mm, clip]{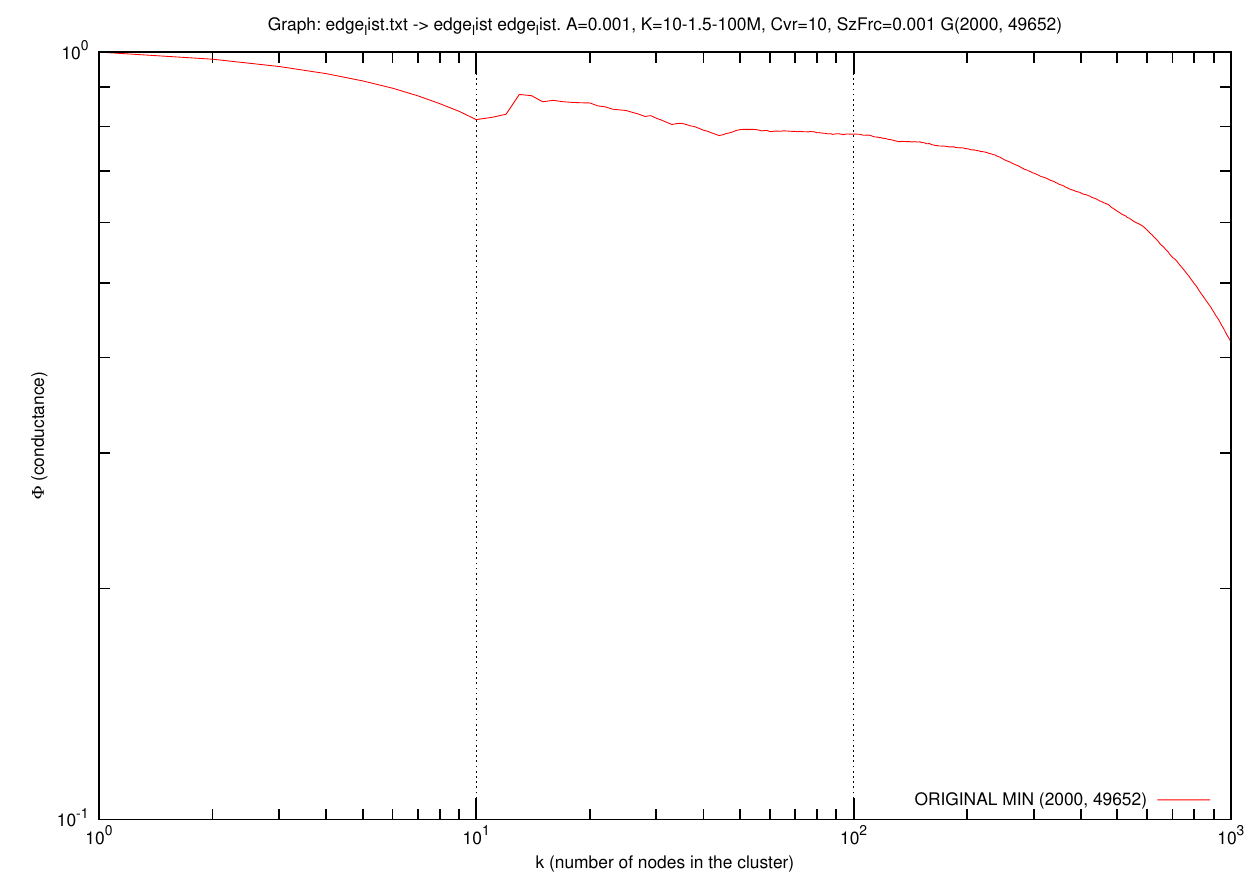}
\caption{ A \emph{network community profile} \cite{LeskovecNCP}, of an LFR graph suggesting that \emph{conductance} is high for all subgraphs of this synthetic graph.
The parameters of LFR synthetic benchmark graph are: 
\texttt{benchmark -N2000 -k50 -maxk50 -minc10 -maxc10 -t10 -t20 -mu0.1 -on2000 -om5}.
Each node is in exactly five communities and \alg{} can detect this highly overlapping structure. }
\label{NCP}
\end{figure}

Having described the model and the corresponding objective function, and the algorithm we propose, there are a number of experiments we performed.
Some of these experiments tell us about the suitability of the objective function, others tell us about the performance of the algorithm.
Firstly, we discuss a question which is not specific to \alg{}. Namely, whether or not a typical empirical graph has highly overlapping structure,
and whether or not an algorithm could ever exist to reliably detect that structure.

A cluster of nodes with low \emph{conductance} with respect to the rest of the graph
can be informally described as a cluster with large internal density
and/or few edges which point out of the cluster.  \cite{LeskovecNCP} analyzed a variety of empirical graphs
and searched for clusters with low \emph{conductance}.
For a given cluster size, $k$, a variety of heuristics are used to search for the single cluster with lowest conductance.
If the conductance values are high for all values of $k$, \cite{LeskovecNCP} argues that this can be interpreted as ruling out community structure at that scale.

In \cref{NCP}, we see the \emph {network community profile} (NCP) plot for a synthetic dataset generated by the
LFR software.
The conductance is greater than $0.4$ for all $k$.
This value is higher than found by \cite{LeskovecNCP}
in a variety of empirical graphs, where values of $0.001$ are not uncommon for some values of $k$.
By the reasoning of \cite{LeskovecNCP}, we might (incorrectly) interpret this as proving that community structure does not exist at any
scale. However, this data was generated with highly overlapping community structure
where each node is placed in five communities. Also, the \Alg is able to detect this structure with high accuracy,
acheiving $92.6\%$ accuracy according to the overlapping extension to NMI.

One relevant hypothesis is that many empirical networks might contain highly overlapping, and easily detectable, community structure
and that such structure may exist at the large scale as well as at the small scale.
\cref{NCP} shows us that large conductance values are compatible with detectable community structure at small scales, and suggests
that large scale structure cannot be ruled out at larger scales.
The hypothesis is not ruled in, but nor can it be ruled out by the NCP.

We do not propose that the \alg\ model is a complete description of how
empirical networks form, nor that the \alg\ algorithm is the only way to detect such structure.
Instead, our aim in this NCP experiment is to show that empirical networks might have
strong, highly overlapping structures at small and large scales, and that
current or future algorithms may be able to reliably detect these communities.


\subsection{Data from the MOSES model}

In many of our experiments, data was generated from the LFR model as
that is becoming quite common in the community finding literature. But we can also
generate data from the MOSES model directly.

As currently formulated, the MOSES model specifies that, \emph{a priori}, the
community sizes are drawn uniformly between zero and $N$, the number of nodes in
the graph. But that is not very realistic, so instead we select community
sizes between 15 and 60 in order to be consistent with the LFR experiments which we will discuss in \cref{sec:AlgoOrModel,sec:LFR}.

We created ten networks, where the average overlap increased from one to 10.
We define the \emph{average overlap} by considering each node and the number of communities
that it is in, then taking the mean.
The average overlap is referred to as $o_m$ in the LFR manual, but we used our own software
for the experiment described in this subsection.
The densities ($p_{in}$ and $p_{out}$) were chosen such that the average degrees would
match those of the networks used in \cref{NMIs} -- average degree $15 \times o_m$ where 20\% of a typical node's edges are not inside any community .
Again, there were 2000 nodes
in each network. To generate the data, we create the required number of communities by selecting nodes randomly,
with replacement, and joining them with probability $p_{in}$. Finally we add extra ``background''
edges between every pair of nodes with probability $p_{out}$.

In \cref{DataFromMOSES} we see that the algorithm can achieve around 85\% NMI, and
a good estimate of the number of communities, up to approximately 10
communities per node. This accuracy can be increased by increasing $p_{in}$
but we chose $p_{in} = 0.33$ as this is where its performance starts to fall on these synthetic
networks, and because it matches the density used in our LFR experiments.

\begin{table}[h]
	\centering
	\caption{Graphs from the MOSES model. $p_{in} = 0.33$, $p_{out} = 0.0015 \times o_m$, $N=2000$}
	\begin{tabular}{|r|r|r|r|l|r|r|}
	\hline
	& \#true             &         & \#found             &     &   \multicolumn{2}{l|}{Estimates from \alg}        \\
	      $o_m$ &        communities & \#edges &         communities & NMI & $\hat{p}_{in}$     & $\hat{p}_{out}$    \\
	\hline
1	& 53	& \numprint{17041}	& 50	& 0.885581	 & 0.228 & 0.00156947 \\
2	& 107	& \numprint{33042}	& 112	& 0.918681	 & 0.316 & 0.00305844 \\
3	& 160	& \numprint{48441}	& 169	& 0.937397	 & 0.323 & 0.00447787 \\
4	& 213	& \numprint{64573}	& 234	& 0.919665	 & 0.334 & 0.00596004 \\
5	& 267	& \numprint{78931}	& 306	& 0.885562	 & 0.335 & 0.00793281 \\
6	& 320	& \numprint{94846}	& 342	& 0.901794	 & 0.336 & 0.00959871 \\
7	& 373	& \numprint{110539}	& 389	& 0.886371	 & 0.336 & 0.0116144 \\
8	& 427	& \numprint{127609}	& 412	& 0.881646	 & 0.336 & 0.0127759 \\
9	& 480	& \numprint{143526}	& 448	& 0.85636	 & 0.335 & 0.0154588 \\
10	& 533	& \numprint{157471}	& 461	& 0.843449	 & 0.336 & 0.0170047 \\
12	& 640	& \numprint{185298}	& 514	& 0.795805	 & 0.338 & 0.0226332 \\
15	& 800	& \numprint{229450}	& 462	& 0.699846	 & 0.337 & 0.0400962 \\
20	& 1067	& \numprint{299020}	& 316	& 0.451244	 & 0.338 & 0.0859497 \\
	\hline
\end{tabular}
\label{DataFromMOSES}
\end{table}

\subsection{The algorithm or the model?}
\label{sec:AlgoOrModel}

\begin{table}[h]
  \centering
  \begin{small}	
    \begin{tabular}{|c|l|c|}
      \hline
      \textbf{Parameter}	& \textbf{Description}  & \;\textbf{Value}\;
      \\ \hline   
        \hline $N$          & number of nodes 		& 2000
	\\\hline $k$          & average degree 		& $15 \times O_m$ 
	\\\hline $k_{max}$    & max degree 	        & $15 \times O_m$ (in \cref{k15mult1})  
	\\                    &                         &  \textbf{or} $45 \times O_m$ (in \cref{k15mult3})
      \\\hline $C_{min}$    & minimum community size 	& 15
      \\\hline $C_{max}$    & maximum community size	& 60 
      \\\hline $\tau_{1}$   & degree exponent 	        & -2 
      \\\hline $\tau_{2}$   & community size exponent 	& 0
      \\\hline $\mu$        & mixing parameter 	        & 0.2 
      \\\hline $O_{n}$      & overlapping nodes 	& 0 \ldots $N$
      \\\hline $O_{m}$      & communities per node 	& 1, 1.2, 1.4, \ldots, 2.0, 3.0, \ldots 10
      \\ \hline   
    \end{tabular}
  \end{small}
  \caption{Parameter values used for the experiments described in \cref{sec:AlgoOrModel,sec:LFR}. Where $O_m = 1.4$, for example, we put 40\% of the nodes, i.e. 800 of them, each into two communities. }
  \label{tab:params}
\end{table}

As discussed in \cref{sec:methods} the \Alg\  is a heuristic optimisation strategy, targeted at finding the set of communities that maximises the
posterior density $\mathrm{P}(\mathrm{z}, p_{in}, p_o | x)$ under our proposed stochastic model.
We have seen good performance on a number of synthetic benchmarks, but it is worth asking, whenever \alg\ fails, is this due to a failure of the heuristic optimisation strategy to find a good fit to the model, or is this due to a failure of the model to properly capture the characteristics of the underlying community structure. 
To investigate this, we looked again at the experiments where the performance of \alg\ breaks down, at 10 or more communities-per-node.
In the case where there are no communities in $\mathbf{z}$, the \alg\ model is identical to the Erdos-Renyi
model. If we optimize $p_{out} = \frac{2 \#edges}{N \times (N-1)}$ then we can use this model as a ``baseline'' value
for the objective function. Then, the ratio of the logs of this quantity to  $P(z, p_{out}, p_{in} | x)$ gives
a value between $0$ and $1$. The \Alg\ attempts to minimize this quantity.

$$
f(z) = \frac  { \log \mathrm{P}_\text{moses} (z, \hat{p_i}, \hat{p_o} | x) }
              { \log \mathrm{P}_\text{Erdos-Renyi} (x | \hat{p}) } 
$$
where $\hat{p_i}, \hat{p_o}$ are optimized for $z$ under the \alg{} model.

\begin{table}[h]
	\centering
	\caption{\label{table:algormodel} What does the \Alg\ target? }
	\begin{tabular}{|r|r|r|r|r|}
		\hline
		Overlap & f($z_\text{ground truth}$) & f($z_\text{moses}$) & $\Delta f$ & NMI \\
		\hline
1	& 0.640671	& 0.641113	&  0.000442	& 0.813046 \\
2	& 0.704139	& 0.715620	&  0.011481	& 0.735667 \\
3	& 0.735461	& 0.738797	&  0.003336	& 0.721188 \\
4	& 0.785880	& 0.793339	&  0.007459	& 0.664248 \\
5	& 0.810475	& 0.814468	&  0.003993	& 0.638435 \\
6	& 0.832514	& 0.829670	& -0.002844	& 0.613011 \\
7	& 0.849016	& 0.841034	& -0.007982	& 0.600076 \\
8	& 0.861816	& 0.848052	& -0.013764	& 0.600137 \\
9	& 0.882812	& 0.864082	& -0.018730	& 0.561496 \\
10	& 0.895454	& 0.870607	& -0.024847	& 0.547616 \\
		\hline
	\end{tabular}
	\label{AlgoOrModel}
\end{table}

In \cref{table:algormodel}, the value of $f()$ is computed for the communities found by the \Alg\ and also for the ground truth communities. 
In all cases, the difference in $f()$ is relatively small, suggesting that the \Alg\ has found communities which are of as good quality as the ground truth communities, according to the \alg\ model.
At the end of \cref{AlgoOrModel}, among the most highly-overlapping datasets, 
we see that the \Alg\ is achieving values of $f()$ which are slightly better than that of the ground truth. This suggests that
although the \alg\ model is not the ideal model for these datasets, the \Alg\  is quite effective at targeting communities that fit the \alg\ model
when the amount of overlap is high.
In order to improve the overall results (the NMI column in \cref{AlgoOrModel}), it will likely be necessary to consider a new model.

It should be noted that the \Alg is a generic algorithm and its heuristics are not restricted to the \alg\ model.
Hence the algorithm could perhaps be applied to other objectives.

\subsection{Evaluation on benchmark data with variable overlap}
\label{InHomogenous}

\begin{figure}[htbp]
\centering
\includegraphics[scale=0.75,trim = 0mm 5mm 0mm 5mm, clip]{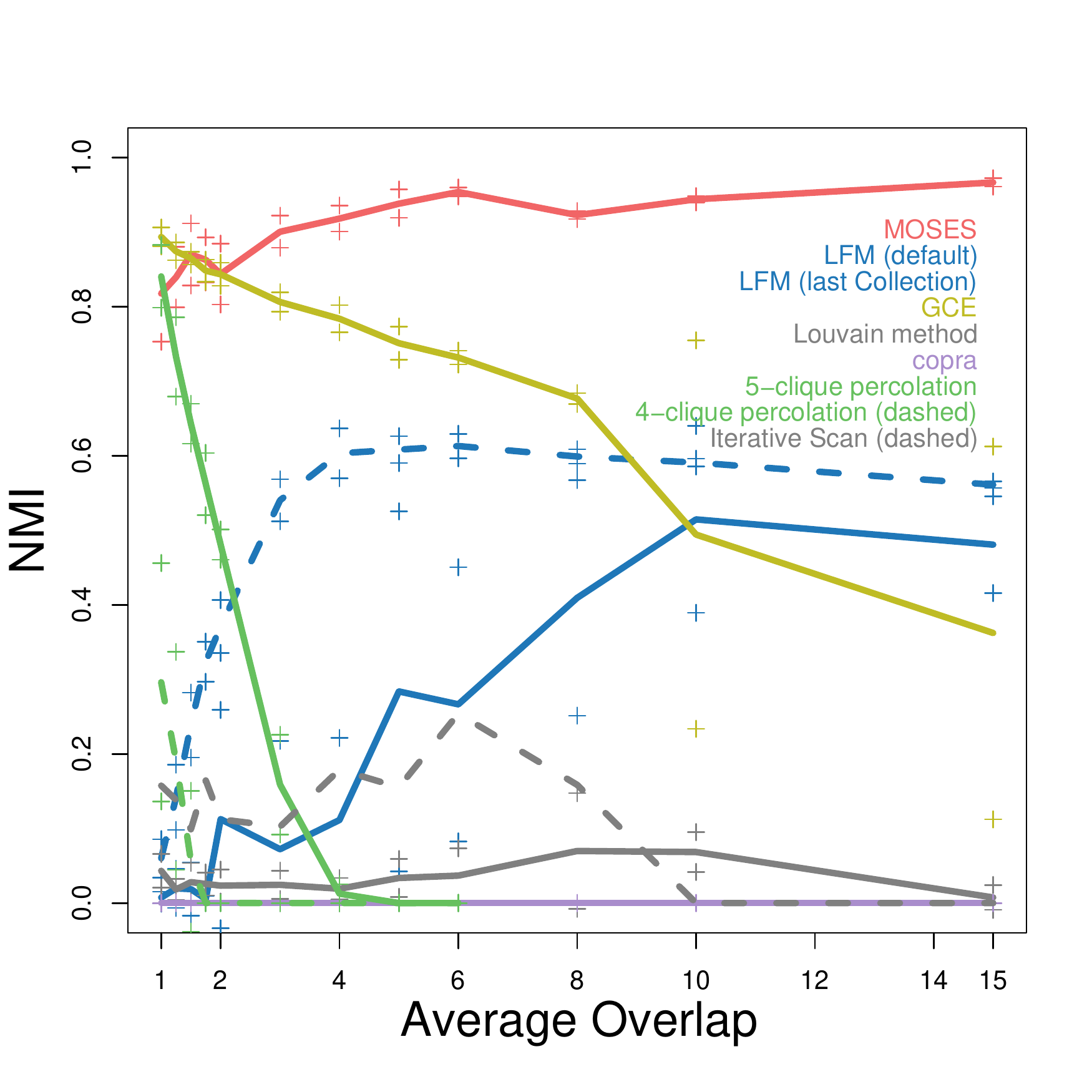} 
\caption{NMI of various algorithms as average overlap increases. Mean +/- standard deviation of twenty realizations of the graph. Iterative Scan (\cite{baumes-2005}) was run just once, due to time constraints.}
\label{NMIm20i1.0}
\end{figure}

To evaluate the accuracy of \alg{} and other algorithms, we created a set of simple
benchmark graphs with increasing levels of overlap.
To generate the graphs, we begin with a graph with 2,000 nodes and no edges. We then assign a number of communities. For each community, 20 nodes are selected at random, without regard to whether those nodes have already been assigned to other communities.  A note on terminology: in this we use \emph{highly overlapping} to mean nodes that are members of many communities.
It is worth considering whether alternative terminology is more appropriate, especially when looking at the intersection of two communities where the intersection may contain many nodes.

We then add edges with reference to these communities. For each community, we add edges until every node is joined to every other node in that community. This gives us a graph with a large number of 20-cliques.

These communities are referred to as the \emph{ground truth communities}. Finally, every pair of nodes is joined with probability 0.005 to add a number of non-community edges. We further confirmed that, in our evaluation, all graphs generated are connected graphs, even those with the smallest number of communities.

We then apply MOSES, and other algorithms, to these graphs to find communities.  We use a recently published extension of
normalized mutual information (NMI) to calculate how similar the
ground truth communities are to the communities found by the various algorithms, as this measure has been popular in the recent literature.
\footnote{For creating the LFR graphs with fixed overlap-per-node and
measuring overlapping NMI, we use the implementations provided by the authors, both of 
which are freely available at \url{http://sites.google.com/site/andrealancichinetti/software}.
For the specification of overlapping NMI, see the appendix of \citet{lancichinetti-2009}.}

Results on these synthetic graphs are shown in \cref{NMIm20i1.0}.  We plot the accuracy, as measured by NMI, of a variety of overlapping
CAAs. On the horizontal axis, we plot the average \emph{overlap}, or average number of communities that a single node is in, within the benchmark graph.
For example, where the average overlap is 1.0, this means there were 100 communities, each of
20 nodes, placed in the 2,000-node graph.

The algorithms used are LFM by \cite{lancichinetti-2009}, COPRA by \cite{gregory-copra}
, Iterative Scan (IS) by \cite{baumes-2005}
, clique percolation, and GCE by \cite{GCE}.
We include the Louvain method \cite{blondel-2008} as an example of a popular partitioning algorithm.
We have used implementations supplied by the authors, except for clique percolation. For clique percolation,  we used our own implementation 
as existing implementations by \cite{SCP} and \cite{palla-2005} were slow on many of the datasets.
The LFM community finding algorithm, and the LFR synthetic network creation software, are not to be confused with each other
but they do share authors.
The  LFM software creates many complete collections from a graph, each of which is a complete community assignment.
As recommended by the authors, we select the first such community assignment for use in this comparison.
However, we have noticed that the results obtained from LFM when selecting the last collection, instead of the first, can be
better. For completeness, we have included this in our comparison.

\begin{figure}[h]
\centering
\includegraphics[scale=0.50,trim = 0mm 5mm 0mm 10mm,clip]{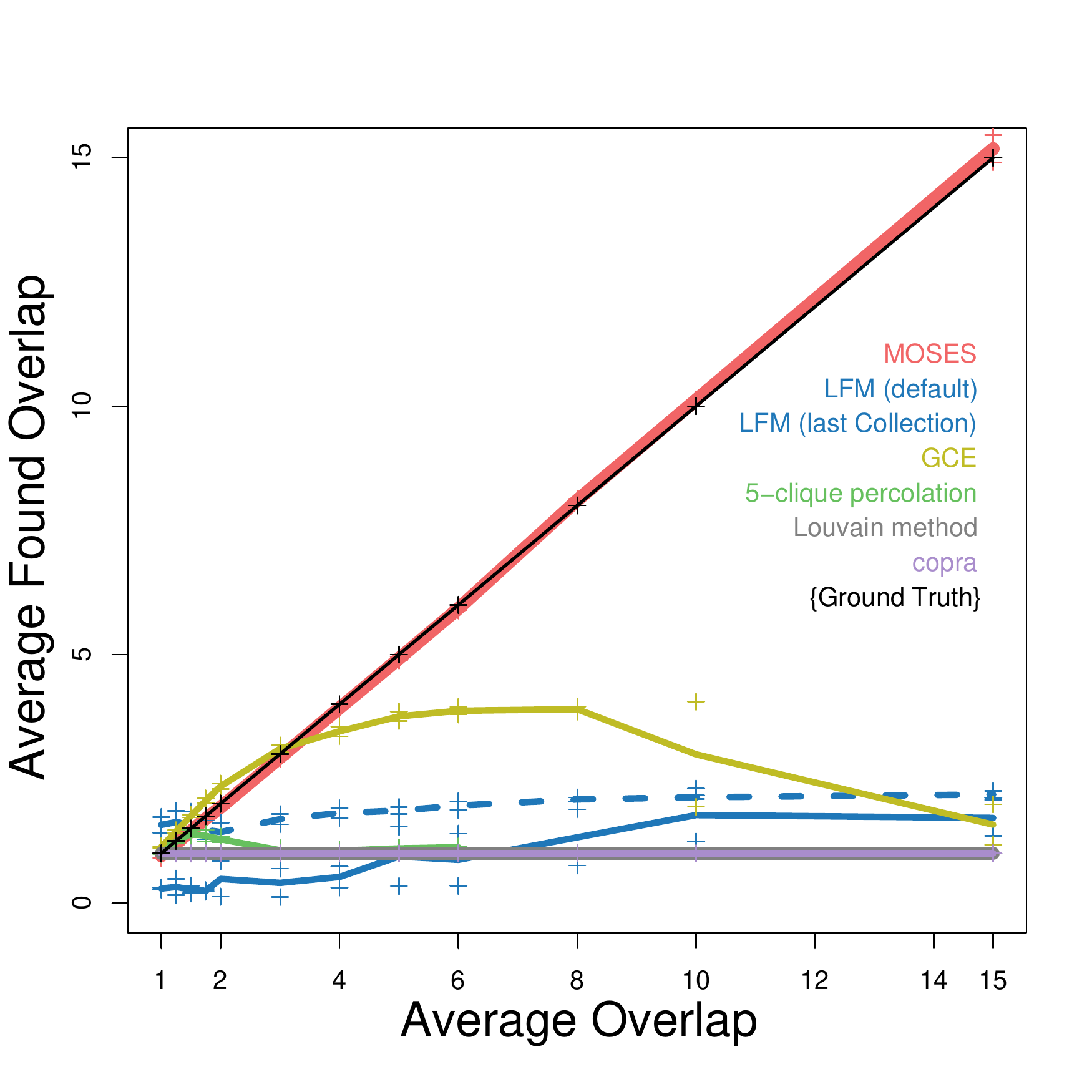} 
\caption{Estimated overlap of various algorithms as average overlap increases.}
\label{AvgOverlapm20i1.0}
\end{figure}

In \cref{AvgOverlapm20i1.0}, we plot the average overlap found by the various algorithms. Only \alg{} is able 
to obtain good estimates of the average overlap, up to an average overlap of 15 communities-per-node.



\begin{figure}
\centering
\subfigure{  \includegraphics[scale=0.28]{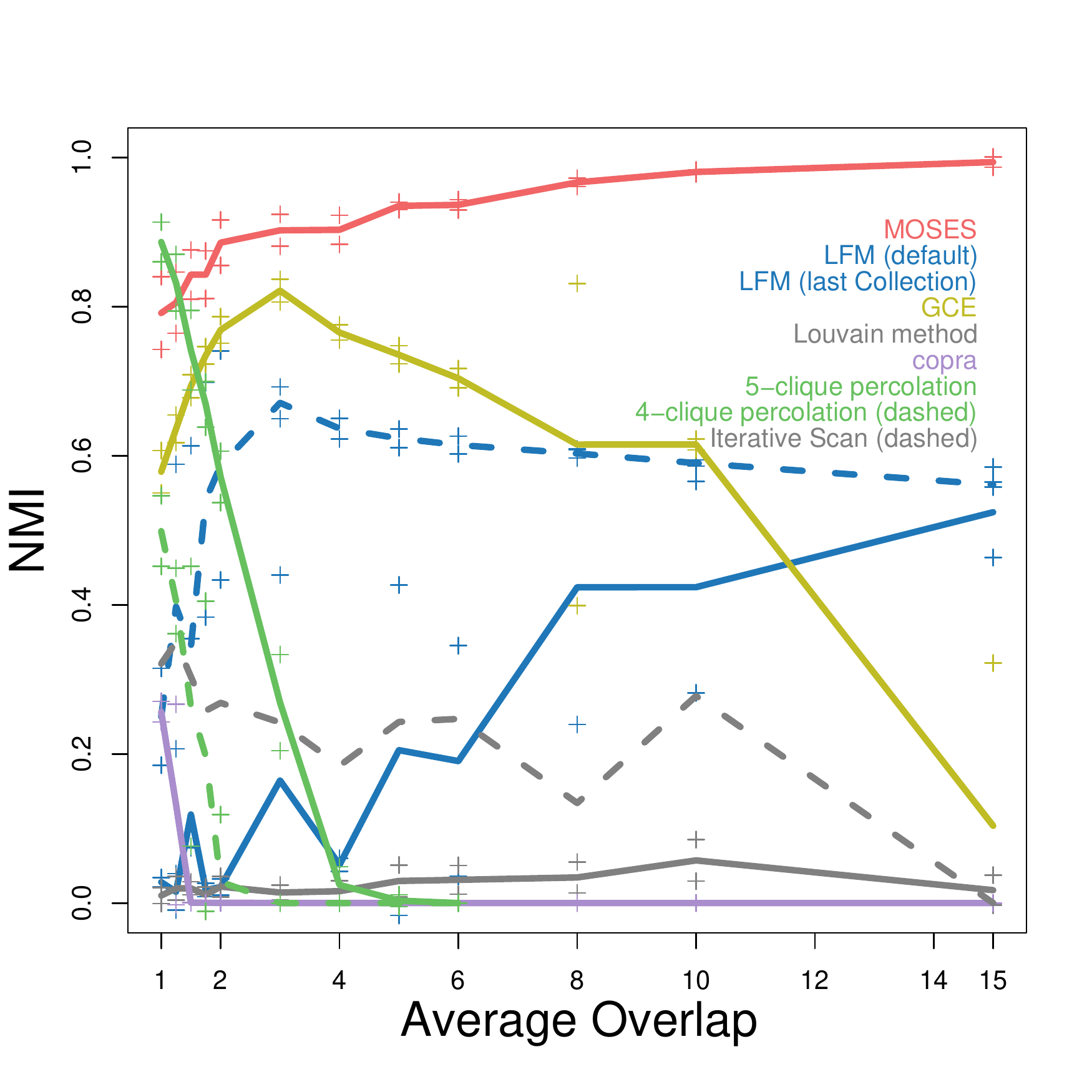}  }
\subfigure{  \includegraphics[scale=0.28]{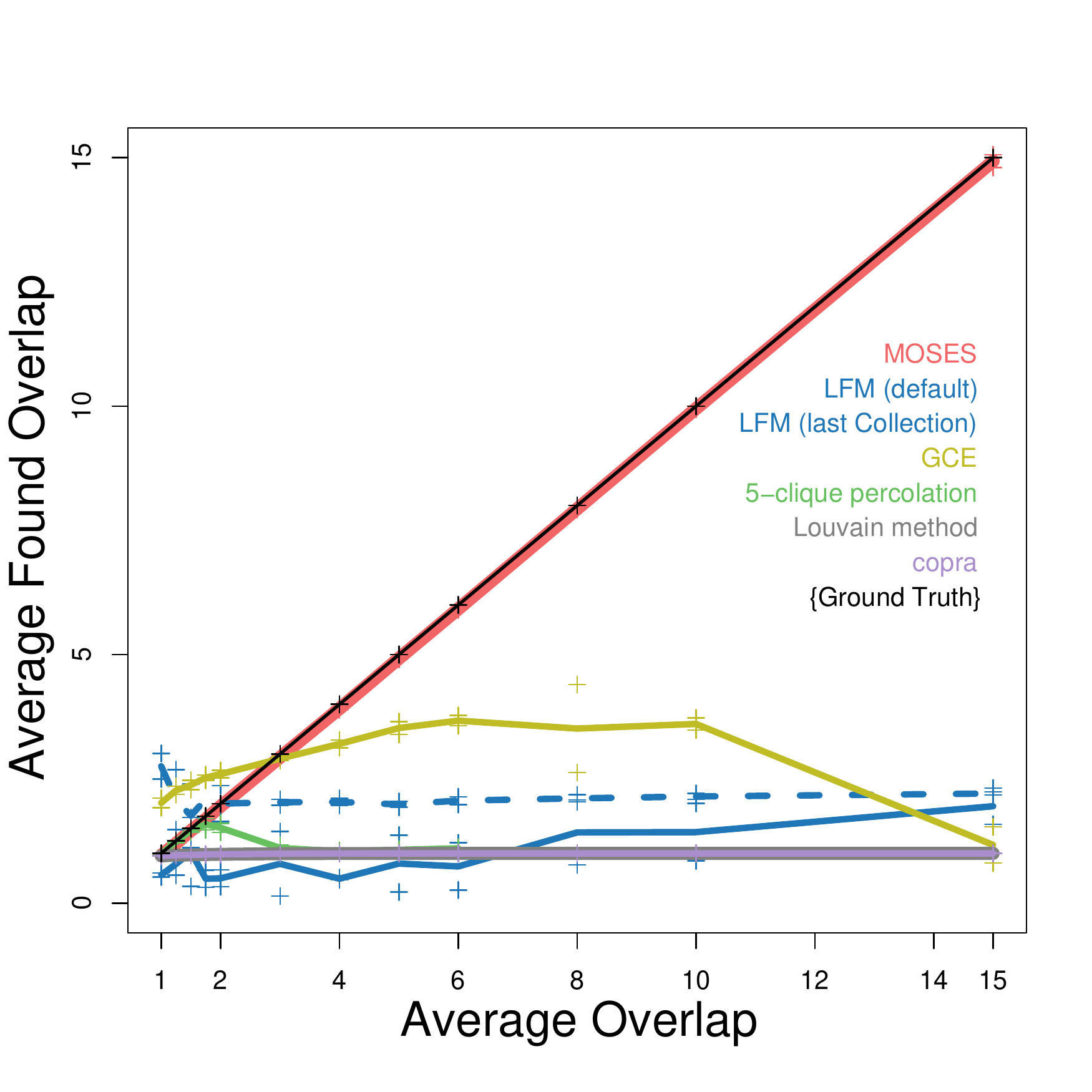}  }
\caption{Graphs with lower levels of ``background'' edges. }
\label{ReallySparseNotEvenConnected}
\end{figure}

In \cref{ReallySparseNotEvenConnected}, we consider graphs with a lower probability, 0.001, for the probability
of a non-community edge between two nodes. This will assign approximately two non-community edges, on average, to each node.
This improves the performance of many algorithms as the number of noisy edges has significantly decreased.
We should note that these graphs are not necessarily connected, and some algorithms operate only on the largest connected
component. For each of these sparser graphs, at least 90\% of the nodes are in the largest connected component.

\begin{figure}
\centering
\subfigure  {  \includegraphics[scale=0.28,trim = 0mm 5mm 0mm 20mm, clip]{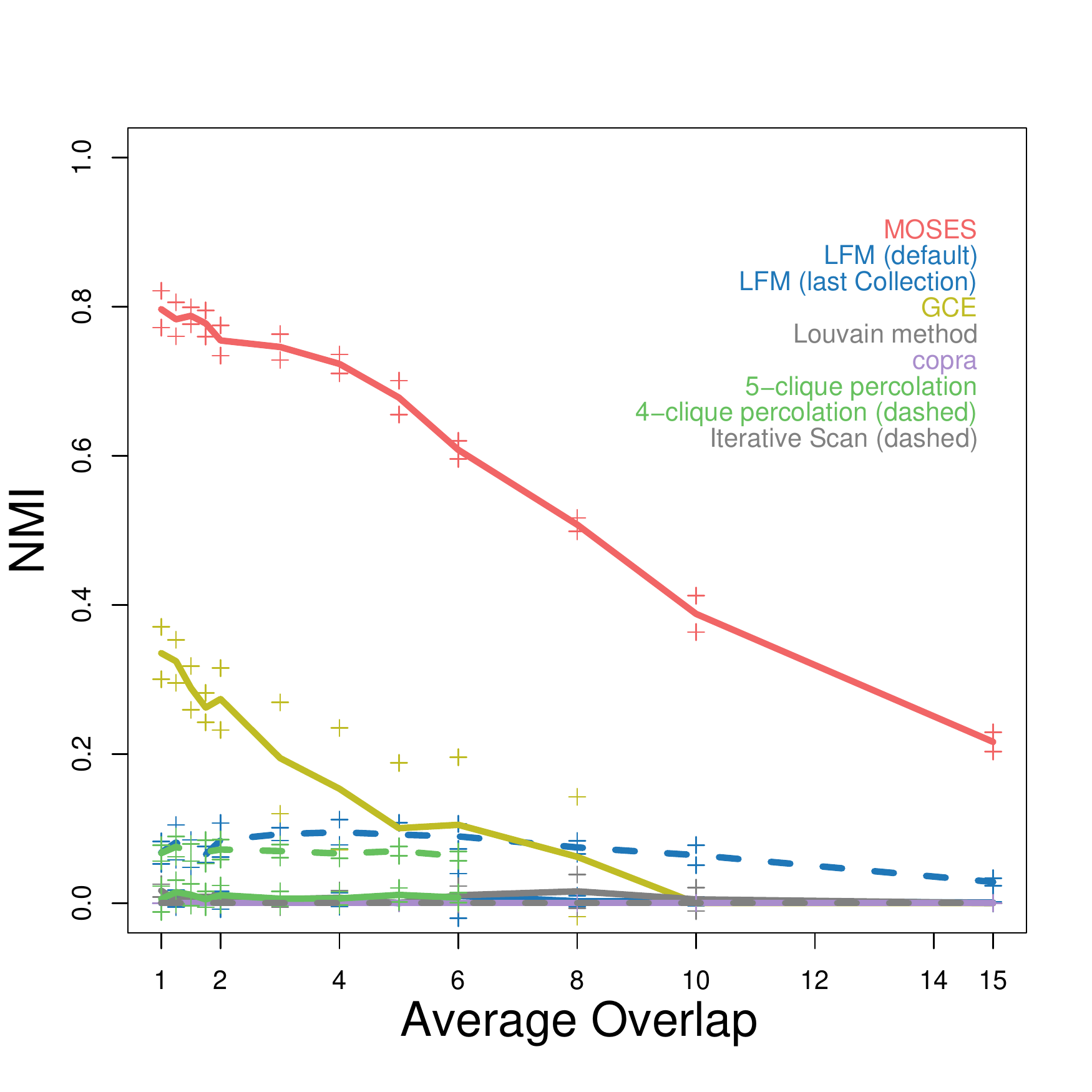}  }
\subfigure  {\includegraphics[scale=0.28,trim = 0mm 5mm 0mm 20mm, clip]{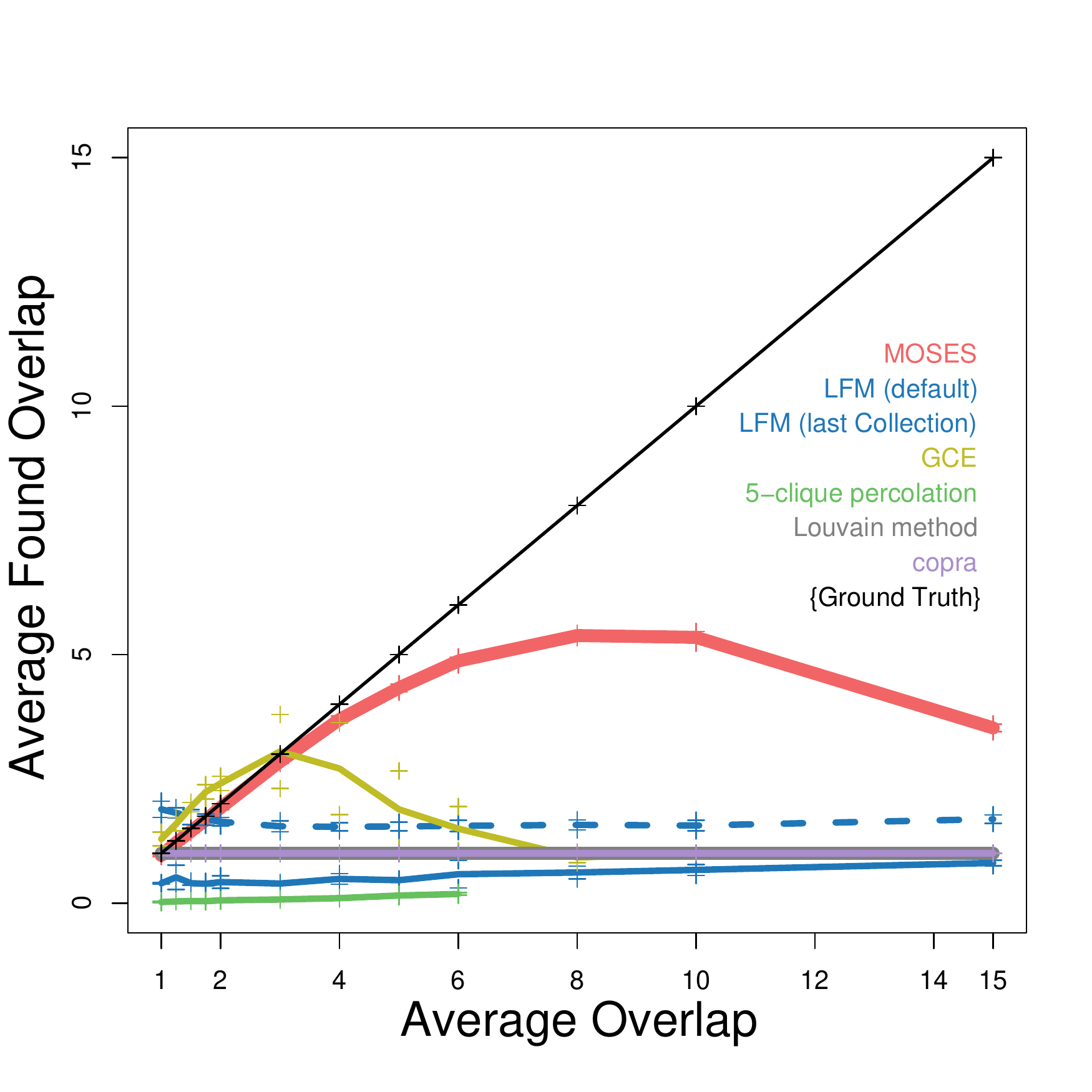}  }
\caption{NMI, and average found overlap, for $p_{in}$ = 0.3. $p_o$ = 0.005.}
\label{ZeroPointThreeAndFour}
\end{figure}

In the benchmarks described so far, each community was a clique, rendering it simple for \alg{} to detect. To investigate
further, we generated a series of benchmarks where the edges inside communities are connected with a lower probability.
As expected, the performance of all algorithms dropped as the internal edge
density decreased. These are presented in \cref{ZeroPointThreeAndFour}.
\alg{} can detect communities at up to 15 communities per node, even as $p_{in}$ drops 
below 0.4. At $p_{in}$=0.3 however, all algorithms tested, including \alg{}, have poor performance.

Considering the broader implications of these experiments, especially \cref{AvgOverlapm20i1.0}, we see that
existing algorithms may underestimate the number of communities. This echoes our earlier hypothesis, that
many empirical networks may have very highly overlapping community structure, which is missed by existing
algorithms.


We also note that the synthetic graphs just discussed are particularly suited to the MOSES model, as all the
communities are created with the same edge density. This homogenous edge density across all communities is a good
match for the $p_{in}$ parameter. In order to investigate performance where the density varies across the ground
truth communities, we next look at LFR benchmark graphs. 


\subsection{Evaluation on LFR Graphs}
\label{sec:LFR}

The LFR benchmark generation software \cite{fortunato-2008} can be used to generate more interesting datasets than those just analyzed.
Above, we looked at communities of a fixed size with a constant internal edge density. The LFR software can generate graphs
with a variety in community size, and a variety in node degree, each of which will create variance in the internal edge density.
Such variety of density provides a challenge to \alg{}, as the $p_{in}$ parameter in the \alg{} model is such that all communities
are assumed to be equally dense.

One drawback of the LFR graphs is that all the overlapping nodes must be assigned to the same number of communities.
This is why we created our own benchmarks , with varying overlap, in \cref{InHomogenous}.
We used the LFR software to generate graphs not unlike those analyzed in the last section. The number of nodes
is again 2,000. The community sizes range uniformly from 15 to 60. The mixing parameter, $\mu$, is 0.2 meaning
that 80\% of the edges are between nodes that share a community.

\begin{figure}
\centering
\subfigure[Fixed degree. $\bar k=15 \times \text{overlap} $] { \includegraphics[scale=0.23,trim = 0mm 6mm 0mm 10mm,clip]{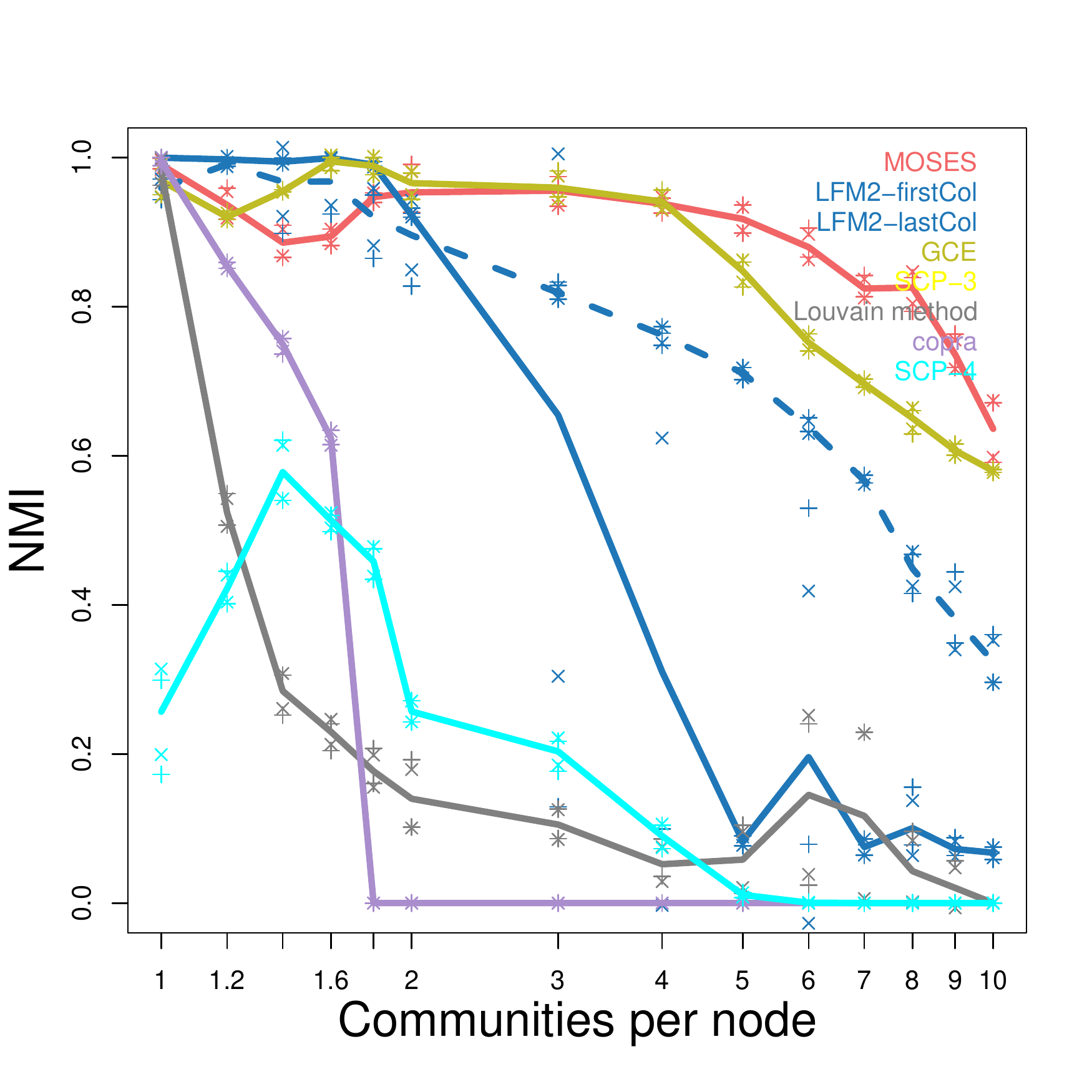} \label{k15mult1} }
\subfigure[Maximum degree is triple average] { \includegraphics[scale=0.23,trim = 0mm 6mm 0mm 10mm,clip]{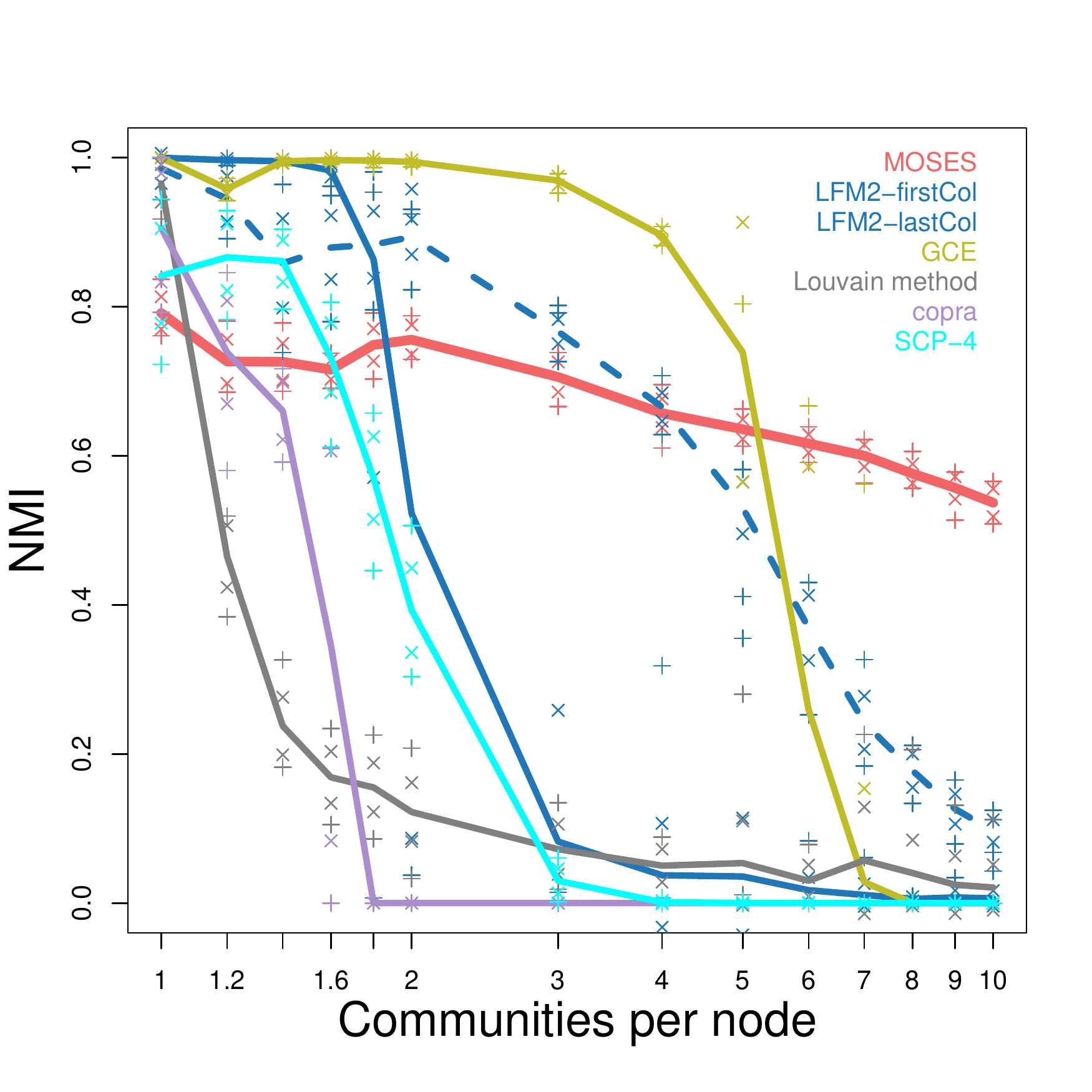} \label{k15mult3} }
\caption{NMI scores as the amount of overlap increases in the LFR fixed-overlap graphs. We mark the mean +/- standard deviation, along with lines through the mean, over twenty realizations of the synthetic benchmark.}
\label{NMIs}
\end{figure}

We varied the overlap to range from one community per node to ten communities per node. Then the degree of all the nodes was fixed
to be 15 times the overlap. We present these results in \cref{k15mult1}, where the horizontal axis is logarithmic.
LFR can create graphs where only a portion of the nodes are assigned to more than one community, we use this feature
to investigate graphs with on average $1.2$, $1.4$, $1.6$, $1.8$ communities-per-node.
Our parameters are summarized in \cref{tab:params}.

In any one of these graphs, each overlapping node is in exactly the same number of communities, making the structure relatively simple.
It is not surprising therefore that many algorithms, such as LFM and the partitioning algorithm by \citet{blondel-2008}, perform well when the overlap is low.

In the previous section we saw that a partitioning algorithm, such as \cite{blondel-2008}, can fail
on graphs with low \emph{average} levels of overlap.
This demonstrates that, even in empirical graphs where overlapping communities are not expected to be major feature, it
may not be wise to use a partitioning algorithm. Partitioning algorithms might succeed only where each node is
known to be in \emph{exactly} one community. This is an unrealistic assumption in many empirical datasets.

The LFR software can generate networks with a power law degree sequence. 
In \cref{k15mult3} we analyzed the same datasets as in \cref{k15mult1} but where the maximum degree
was set to be three times the average degree. The slope of the power-law is
 set it to $2.0$.
In these datasets, when the overlap is low, \alg{} does not perform as well as GCE, LFM or clique percolation.
On the other hand, \alg{} is the only algorithm capable of detecting significant
structure when the overlap approaches 10 communities per node. The NMI of the community assignments found by \alg{}
is consistently above 60\% whereas the other algorithms' scores are well below 40\% when there are more
than six communities per node.

The \alg{} model does not explicitly model degree distribution,
nor does it explicitly model  different within-community densities for different communities and this may explain
its failure to get the highest NMI scores in \cref{k15mult3}. This may be an area for future development of these models.
The superior community model of \alg{} enables it to detect some structure in the graphs with heaviest overlap.

\subsection{Scalability}

\begin{figure}
\centering
\subfigure[Fixed degree. $\bar k=15 \times \text{overlap} $] { \includegraphics[scale=0.30,trim = 0mm 6mm 0mm 10mm,clip]{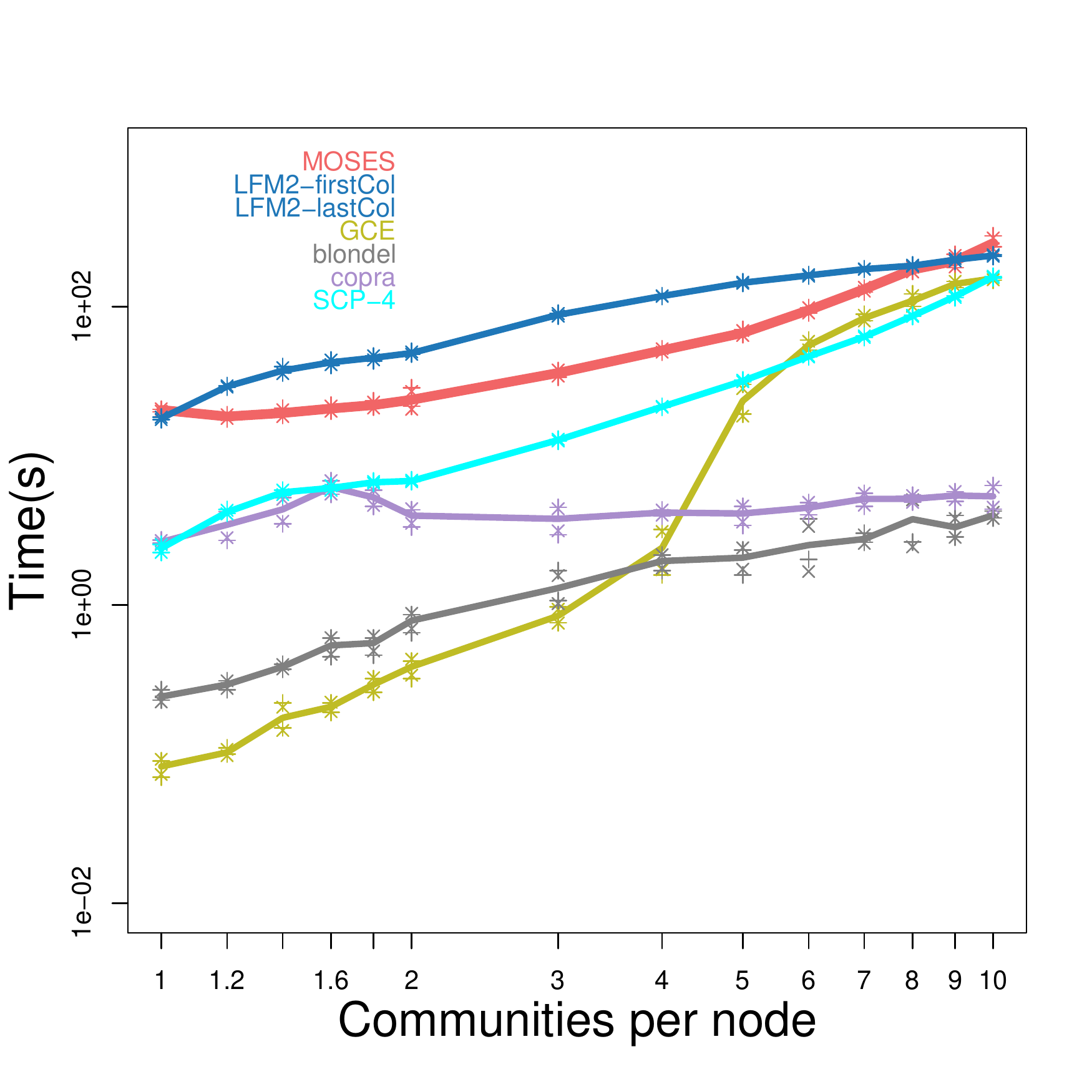} \label{sk15mult1} }
\subfigure[Maximum degree is triple average] { \includegraphics[scale=0.30,trim = 0mm 6mm 0mm 10mm,clip]{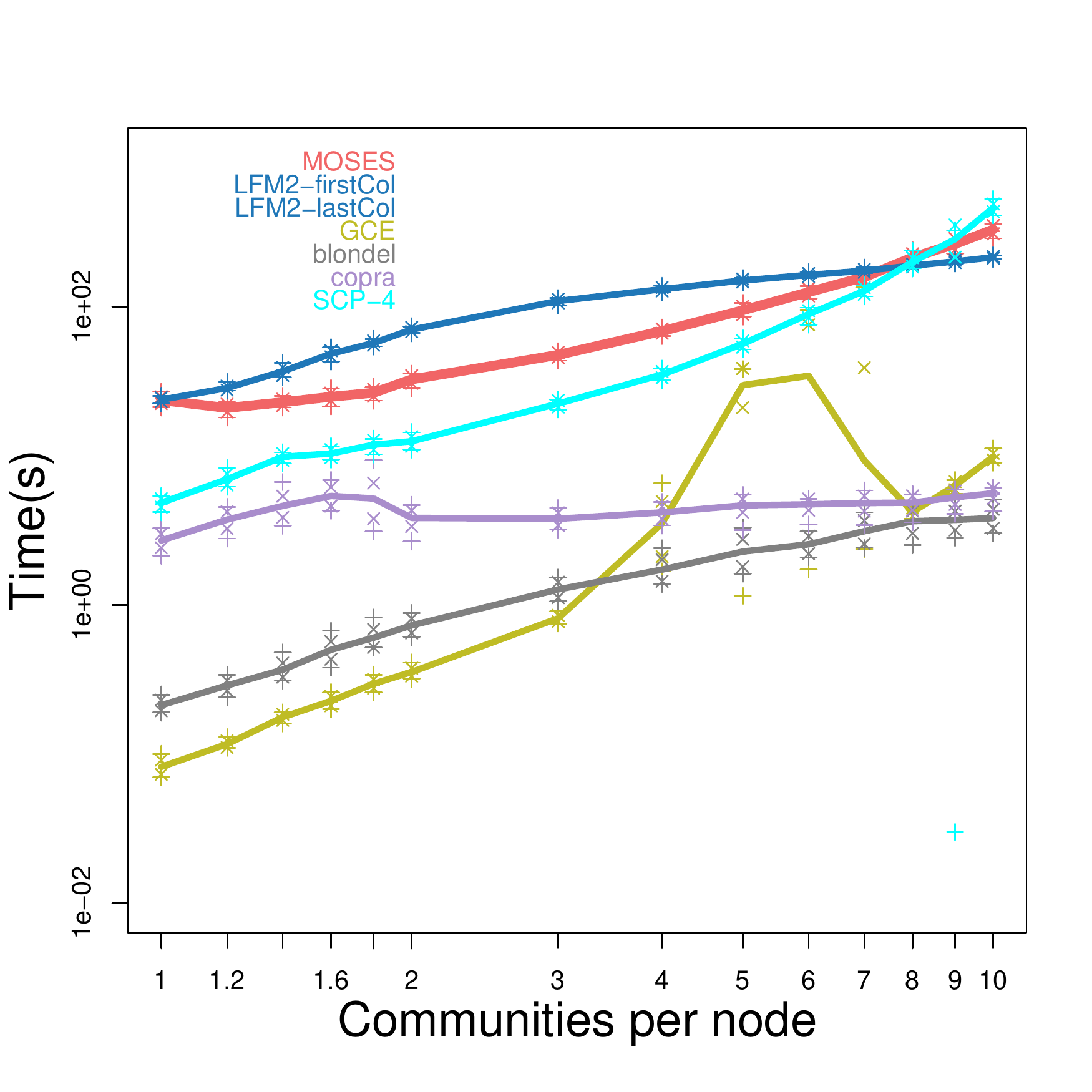} \label{sk15mult3} }
\caption{ Run time, in seconds, as overlap increases in the LFR benchmarks. }
\label{Timings}
\end{figure}

In \cref{Timings} we investigated the run time of these algorithms. The graphs are the same as in \cref{NMIs}, but instead
we plot the logarithm of the running time on the y-axis. GCE is the fastest of all the algorithms on the less overlapping
data. While there are many algorithms faster than \alg{} and LFM,
the only one of those algorithms capable of getting reasonable NMI scores is GCE.
The high quality NMI scores of \alg{} do not carry a significant penalty in performance. \alg{} is as fast as many scalable algorithms
on overlapping data, and gets the highest quality results on the very highly overlapping data.

In partitioning, the most popular and scalable methods can be trivially applied to a variety of objective functions.
The Louvain method, and variants, have been used for both modularity maximization and to maximize the map equation of \cite{TheMapEquation}.
It might be best to think of the Louvain method not as a modularity maximization algorithm, but as a fast method
to maximize any simple partitioning objective function.

For overlapping community finding, we hope to see progress on such ``multi-objective'' algorithms in future.
The MOSES algorithm is not restricted to the MOSES model. And it is also valid to consider very scalable algorithms
which are not based on the MOSES algorithm, but which do target the MOSES model.

\subsection{Evaluation on a real-world social network.}
\label{sec:fb}
\citet{traud-2009} gathered data on Facebook users and friendships in five US universities.

\begin{figure}[]
\centering
\subfigure[Community size] {
	\label{CommSizes}
	\includegraphics[scale=0.18,trim = 0mm 6mm 0mm 20mm,clip]{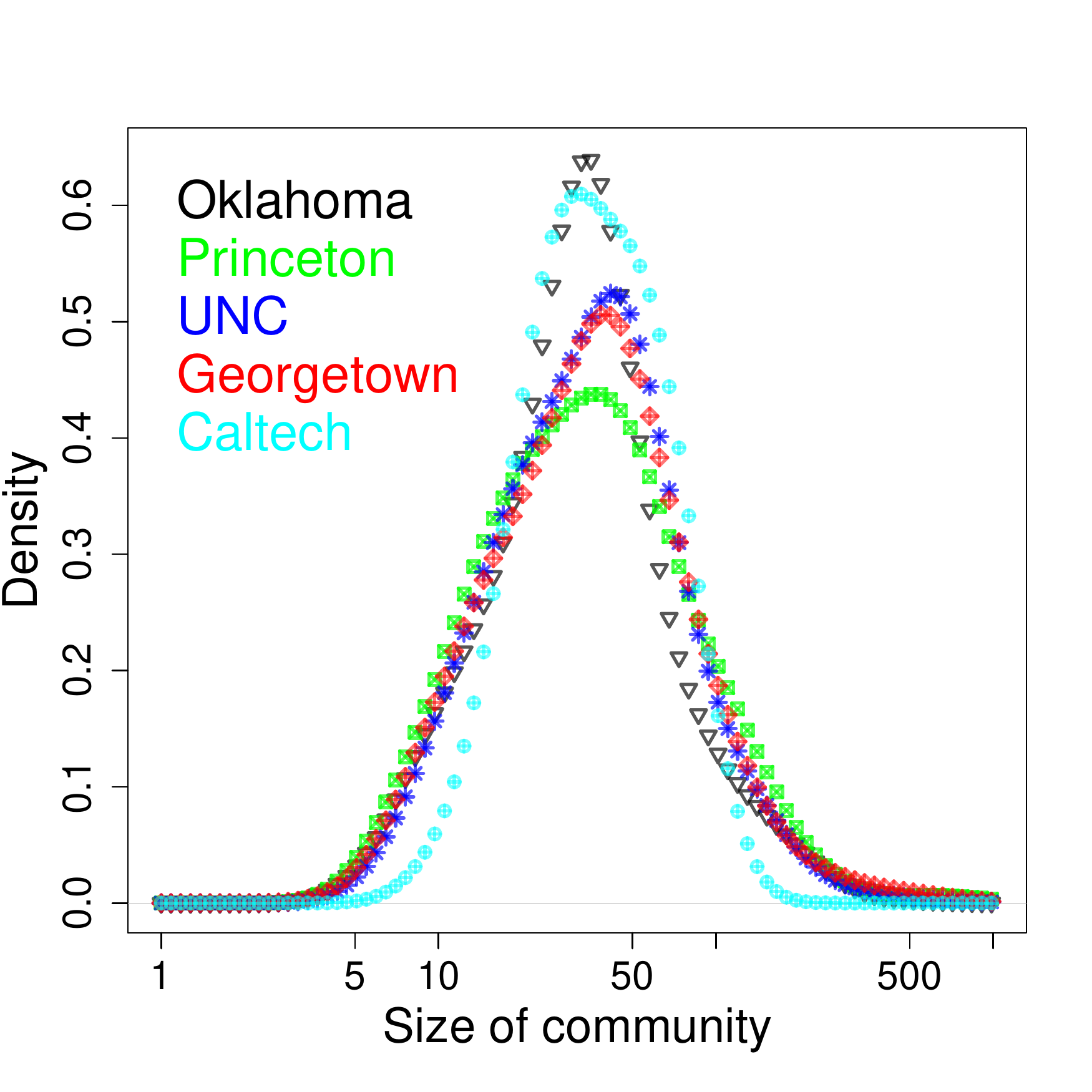} 
}
\subfigure[Degree distribution] {
	\label{DegreeHist}
	\includegraphics[scale=0.18,trim = 0mm 6mm 0mm 20mm,clip]{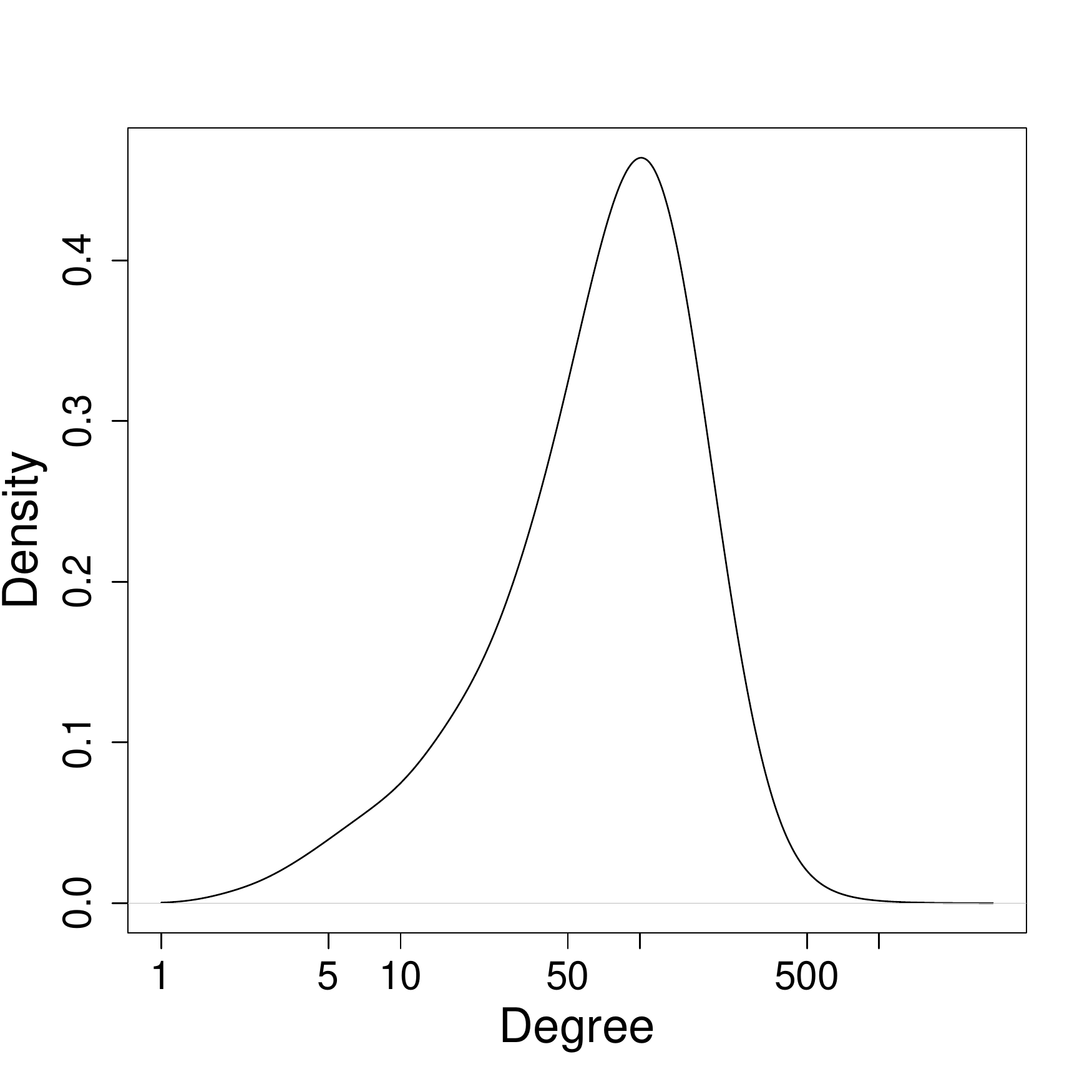}
}
\caption{Sizes of the communities found, and degree distribution for Georgetown, in (logarithmic) density plots.
}
\end{figure}

The degree distributions of all five appears to be very approximately log-normal, as can be seen in the
logarithmic histograms of \cref{DegreeHist}. The distribution does not fit the power law distributions
often assumed as an approximation for the degree distribution of empirical graphs. The relative
narrowness of this (logged) degree distribution may improve the results of \alg{} as it is
a more reasonable fit for the \alg{} model than a strict power law distribution would be. 
The average degree ranges from $43.3$ to $102.4$ across the five universities. Assuming that communities are not very large, and that
most edges in these networks are community edges, it must be the case that the average node is in many
communities.

A summary of the results is presented in \cref{FacebookSummary}. It suggests that
a Facebook user is, on average, a member of seven communities. In an analysis of one of their own 
Facebook ego-networks, \citet{salter2009variational} found it divided into six groups.
\alg{} assigns nodes each to a different number of communities, and to communities of varying size.
In \cref{Georgetown4240}, we present the communities of a student at Georgetown. \alg{} assigns
this student to four communities, and we visualized the subgraph based on all the nodes in those four
communities.


\begin{table}[]
\centering
\caption{\label{FacebookSummary} Summary of \citet{traud-2009}'s five university Facebook datasets, and of \alg{}'s output.}
\begin{tabular}{|r|r|r|r|r|r|}
	 \hline
                     & \boldrotate{Caltech}    & \boldrotate{Princeton}  & \boldrotate{Georgetown \,} & \boldrotate{UNC} & \boldrotate{Oklahoma} \\
	 \hline
 \textbf{Edges}      & 16656      &    293320  &     425638 & 766800 & 892528  \\
 \textbf{Nodes}      &     769    &      6596  &       9414 &  18163 & 17425   \\
 \textbf{Average Degree}     &   43.3     &      88.9    &     90.4 & 84.4  & 102.4    \\
	 \hline
 \textbf{Communities found}      & 62         &       832  &       1284 & 2725  &  3073    \\
 \textbf{Average Overlap}    & 3.29       &      6.28   &    6.67   & 6.96  &  7.46    \\
	 \hline
 \textbf{\alg{} runtime (s)}    & 41 & 553    & 839      & 1585    & 2233     \\
 \textbf{GCE runtime (s)}       & 1  & 1067   & 1657     & 3204    & 664      \\
 \textbf{LFM runtime (s)}       & 23 & 740    & 1359     & 4414    & 4482     \\
	 \hline
\end{tabular}

\end{table}

\begin{figure}[bp]
\centering
\end{figure}

\begin{figure}[tp]
\end{figure}

\pagebreak
\section{Conclusions}
\label{sec:conclusions}

\alg{} detects overlapping community structure
in large networks where nodes may belong to many communities.
Existing algorithms find only relatively low levels of overlapping community structure.
It is necessary to be able to detect highly overlapping structure, if only to rule it out for a given
observed network.
For instance, our analysis on Facebook data has shown that a typical Facebook user can be a member of seven communities.
This demonstrates the need for further research into such community structure.
Existing algorithms work best where each node is in the same number of communities. But this is not
a realistic assumption for social networks and we have demonstrated that \alg{} can accurately
detect communities in networks where typical nodes are in many communities, and where there is
variance in the number of communities a node is in.

\begin{acknowledgements}
Thanks to Prof. Brendan Murphy for providing feedback on the MOSES model.
\end{acknowledgements}

\bibliography{snam}
\bibliographystyle{year}

\end{document}